# Physics Informed Neural Network – Estimated Circuit Parameter Adaptive Modulation of DAB

Saikat Dey, *Student Member, IEEE* and Ayan Mallik, *Senior Member, IEEE*

*Abstract-* This article presents the development, implementation, and validation of a loss-optimized and circuit parameter-sensitive triple-phase-shift (TPS) modulation scheme for a dual-active-bridge (DAB) DC-DC converter. The proposed approach dynamically adjusts control parameters based on circuit parameters estimated using a physics-informed neural network (PINN). For this purpose, first, a precise frequency-domain model for the DAB converter is developed, incorporating resistive and capacitive non-idealities, deadtime effects, and switching network losses. This model closely mimics real hardware behavior with a power flow absolute error margin of within 5%. Secondly, such an analytical model is utilized to synthesize overall system loss optimized TPS control variables as polynomial fitted functions of the converter's output voltage, power, and parameters that are subject to uncertainty such as power transfer inductances and device on-state resistances. Following this, combining a data-driven neural net (NN) and analytical DAB physics model, a PINN architecture is constructed with the purpose of estimating DAB circuit parameters with inputs of DC side voltages and currents, and modulation variables. Notably, the data-light PINN is trained with a mixed dataset collected from real hardware and the analytical DAB model. Further, the trained NN is implemented in a Python environment, with serial peripheral interface (SPI) communication established with the DAB controller enabling real-time updates of the TPS modulation variables based on NN-predicted circuit parameters. The proposed NN-in-loop, parameter-sensitive TPS modulation is validated and benchmarked in a 2kW/100kHz GaN-based DAB converter. With an experimentally emulated 30% reduction in series inductance, the converter achieves a peak efficiency of 98.3%, representing an average efficiency improvement of approximately 1.4% compared to conventional non-adaptive TPS modulation.

## I. Introduction

SINCE its introduction in the early 1990s [1], the dual-active-bridge (DAB) topology has garnered significant attention due to its numerous advantages, including bidirectional power flow capability, high-frequency isolation, efficient wide-gain operation, ease of control, high power density, and zero-voltage-switching (ZVS) capability. Today, DAB converters are widely utilized across various applications, including electric vehicles [2]-[3], energy storage systems [4], solid-state transformers [5]-[6], DC microgrids [7], solar microinverters [8-10] etc.

In a DAB converter, two active H-bridges are connected via a high-frequency transformer (HFT) and series-connected inductances. Power flow between the bridges is controlled by adjusting the phase shift between their square-shaped voltages, with the simplest control method being single-phase shift (SPS) modulation [2]. However, SPS has a limited ZVS range and struggles to maintain high efficiency across wide voltage gain and load conditions. To address these limitations, an inner phase shift within one of the H-bridges can be introduced, adding an extra degree of freedom, leading to dual-phase shift (DPS) or extended-phase shift (EPS) modulation. When both inner phase shifts of the two H-bridges are independently controlled, the system operates under triple-phase shift (TPS) modulation. With three degrees of freedom, TPS represents the superset of all DAB modulation techniques.

Numerous studies have explored DAB modulation optimization strategies, with a focus on minimizing conduction losses [11]-[13], switching losses [14]-[16], peak current stress [15]-[18], inductor current RMS [19]-[20], and expanding the ZVS range [14]-[16], [21]-[22]. These existing modulation strategies can generally be classified into three categories based on their hardware implementation:

1. *Closed-form expression-derived optimal TPS variables*: In this method [11-12], [14-17], [19-21], an approximated analytical time-domain DAB model is used to derive optimal phase-shift variables as closed-form functions of the DAB's voltage gain and load levels. These functions are easily implementable in digital controllers and can update control variables based on sensed voltages and currents. However, due to the use of an idealized, lossless DAB model that neglects circuit parasitics, losses, and dead-time effects, the resulting control variables often lead to suboptimal performance [8], [23].

2. *Look-up table (LUT)-based implementation*: To enhance modeling accuracy and thereby improve modulation performance, some works [8], [24] incorporate system non-idealities, such as parasitic resistances, magnetizing inductance, and device parasitics. In this approach, deriving closed-form expressions for loss-optimized TPS variables becomes difficult, so numerically optimized control variables are stored in 2D or 3D LUTs within the digital controller [8], [22]. The main limitation of this method is the trade-off between the size and resolution of the LUT, constrained by the memory available in the microprocessor.

3. *Artificial neural network (ANN)-based derivation*: Recent studies [18], [25] have employed this approach, training an ANN using extensive datasets obtained from hardware measurements or simulation tools. The trained ANN model predicts optimized phase shifts for specific converter operating conditions. However, the accuracy of these optimized control variables is heavily dependent on the accuracy of the DAB converter model, unless the training dataset is derived from real hardware.

All the aforementioned control strategies for the DAB converter assume that the converter's circuit parameters, such as series inductance and parasitic resistances, are known in advance and remain constant throughout operation. However, the value of inductance is closely tied to flux density, which is highly sensitive to various factors such as temperature, air gap, and switching frequency. Similarly, the device on-resistances can change due to operating temperature fluctuations, device degradation, and other environmental factors. Not accounting for such ambient factor dependent parametric variability can severely degrade the performance of optimized modulation schemes and feedforward control. This degradation can manifest as increased current stress on the active bridges, loss of soft-switching capabilities, efficiency reduction, and potential thermal management issues. The root cause is the controller's inability to update its knowledge of the circuit parameter values in real-time, preventing it from adjusting control variables optimally to maintain peak efficiency. Apart from DAB modulation optimization, the parameter estimation task is also critical for component health monitoring [23], incorporating feedforward control [26] and power balancing control [27] in modular DAB converter.

There are three potential methods to estimate the circuit parameters of the DAB converter and eventually integrate these estimates with optimized control synthesis:

1. *Inductance estimation based on sampled HF current*: The HF current flowing through the DAB inductor can be sampled at switching instants throughout a switching period using a high-bandwidth (BW) current sensor. By applying the volt-second law, the inductance value can be approximated from these sampled current values. However, implementing such high-bandwidth sensors is both costly and prone to noise interference. Additionally, this method provides no information regarding the parasitic resistances in the circuit.

2. *Analytical DAB model derived estimation*: Using the approximated, lossless time-domain DAB model [11], power flow can be expressed as a function of the operating voltages, control variables, and lumped series inductance. By performing reverse computation, the series inductance can be mathematically derived based on the known power flow [26-27]. However, due to the presence of losses and non-idealities in real hardware, such estimations may be highly inaccurate.

3. *Artificial Intelligence-based parameter estimation*: With the rapid development of artificial intelligence (AI) in power electronics [28], machine learning (ML) and deep learning (DL) models are increasingly being employed for power converter parameter estimation [23], [29], based on sensed or sampled voltage and current data. For example, the work in [29] estimates all circuit parameters of a buck converter using a time-domain mapping artificial neural network (ANN). Similarly, a genetic algorithm (GA)-integrated back-propagation neural network (BPNN) is used in [23] to estimate circuit parameters of a dual-active-bridge (DAB) converter. However, the ANN in [23] is trained using a mathematical model that does not account for switching losses and limits modulation to SPS only. Furthermore, data-driven ANN models generally require large datasets to map input-output variables accurately, making them less practical without substantial data collection efforts. This highlights the need to integrate physics-based models with data-driven approaches, ensuring the learning process converges towards physically consistent solutions. Physics-informed neural networks (PINNs) offer such an advantage by incorporating physical principles or physics loss into neural network training [18], [30-31]. Though still emerging in converter parameter estimation, there have been promising works [30]-[31]. For example, Zhao et al. [30] propose a PINN model for estimating L-C components, along with their DCR and ESR, in a buck converter, achieving an average prediction error of ±5%. Despite these advancements, several challenges remain: (a) most models are trained using data from simulations or approximated mathematical models, which do not fully replicate real hardware performance; (b) many works focus on simpler converters like buck converters or basic control strategies with only one degree of freedom, such as duty cycle or SPS modulation and the converters are switching at a low frequency of 20kHz and below; and (c) none of the existing works integrate the circuit parameter estimation platform into an efficient real-time interactive control system.

Addressing these research gaps, this work proposes a NN-in-loop, loss-optimized TPS modulation strategy for DAB converters. The TPS modulation variables are derived from a non-ideality and loss-inclusive frequency domain DAB model, avoiding inaccuracies inherent to idealized lossless models. Additionally, a PINN model is developed to estimate the circuit parameters, trained using a limited dataset gathered from both real hardware and the DAB mathematical model. Finally, the trained neural network is integrated into the synthesized DAB modulator, enabling maximum efficiency tracking with parameter adaptability.

The key contributions of this work can be summarized as follows:

**(a)** A precise mathematical model of the non-ideal DAB converter is developed using a multi-harmonic, frequency-domain approach. This model incorporates critical circuit parasitics, including device on-state resistances, capacitor equivalent series resistance (ESR), ac winding resistances, parasitic capacitances of the HFT, and switching devices, as well as the deadtime effect on the power flow calculations. By accurately modeling high-frequency winding currents, bridge voltages, and soft-switching phenomena, the proposed model enables detailed power loss calculations for each element in the DAB power stage. As a result, this DAB model precisely replicates real hardware performance, eliminating the need for traditional simulation-based models.

**b)** Building on the developed analytical DAB model, a constrained multivariate optimization routine is formulated to compute the optimal modulation variables for any given operating voltage, load, and DAB circuit parameters. This optimization program is further extended to generate polynomial-fitted models that can be implemented in the DAB controller, providing optimal duty cycle variables as functions of sampled input/output DC voltages, currents, and key circuit parameters. This contribution enables a circuit-parameter-sensitive, optimized DAB modulation scheme, facilitating maximum efficiency tracking across a wide range of operating conditions, even when circuit parameters vary.

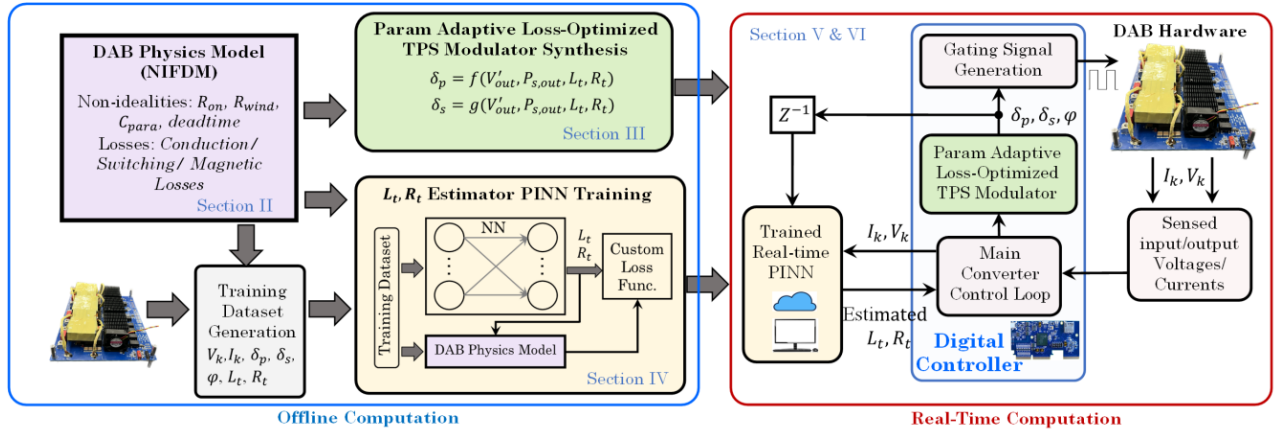

Fig. 1. Proposed strategy to implement a PINN-estimated circuit parameter adaptive overall loss-optimized TPS modulation scheme for a DAB converter. Flowchart highlights the key components of the proposed method and their logical relationship.

(c) A physics-informed neural network (PINN)-based circuit parameter estimation method is proposed. This approach benefits from the inclusion of the analytical DAB loss model in the PINN architecture, enabling faster training with limited data. Once trained, the NN can accurately estimate the DAB series inductances and parasitic resistances based on sampled voltages, currents, and control variables, achieving a mean absolute error of less than 3%.

(d) A NN-in-loop DAB modulation strategy is developed, implemented, and validated, enabling real-time tracking of the maximum efficiency operating point. This approach dynamically adjusts the bridge voltage phase shifts and duty cycles based on estimated circuit parameters, operating voltages, and load conditions, ensuring parameter invariant optimal performance.

To emphasize the key components of this comprehensive work and their logical interrelationships, a flowchart is presented in Fig. 1. This flowchart clearly highlights the major contributions of the study and maps them to their respective sections within the paper. Additionally, it delineates the offline and online computations necessary for implementing the proposed parameter-sensitive modulation scheme.

## II. IMPROVED DAB STEADY STATE MODEL CONSIDERING PARASITIC NON-IDEALITIES AND DEADTIME EFFECTS

The circuit topology of a dc-dc DAB converter with input and output voltages of $V_{in}$ and $V'_{out}$ respectively, is illustrated in Fig 2(a). This configuration consists of two active H-bridges, each using four MOSFETs, connected via a HF transformer with a turns ratio of $n_p:n_s$. The power transfer between the dc ports in a DAB converter, operating at a fixed switching frequency ($f_{sw}$), is controlled by adjusting three key modulation parameters (see Fig. 2(b)): the phase shift ($\varphi$) between the fundamental components of the ac bridge voltages $v_p$ and $v'_s$, and the duty cycles of the bridge voltages, $\delta_p$ and $\delta_s$. In state-of-the-art approaches, under TPS control, these modulation parameters are optimized to minimize transformer winding RMS current, or/and conduction or switching losses. This optimization is typically based on a conventional DAB modeling framework, as outlined below.

### A. Conventional Time-domain Model:

Traditionally DAB converters are analyzed using time domain approach [11-12] with the assumptions of ideal transformer and switching semiconductor devices. The magnetizing inductance $L_m$, parasitic capacitances, winding resistances of the transformer, device on-state resistances are ignored during the power flow and winding current calculations. Further, the presence of parasitic body capacitors in the switching devices and the deadtime effects are ignored during modeling of the Fourier components of the bridge

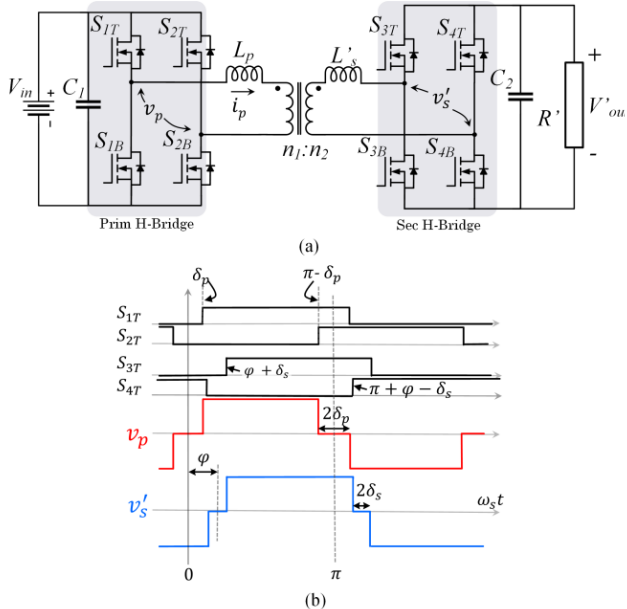

Fig. 2. (a) DAB Converter Topology. (b) H-bridge output voltages $v_p$ and $v'_s$ in relation with the timing diagram of the switching gate signals characterized by $\varphi$, $\delta_p$, and $\delta_s$.

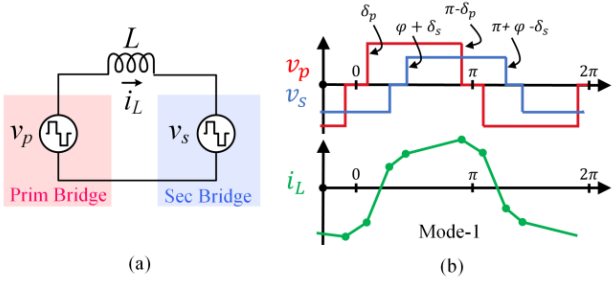

Fig. 3. (a) Primary referred equivalent circuit of DAB using ILM; (b) equivalent model of a DAB; (c)-(g) Switching bridge voltage and inductor current waveforms for a DAB under different operating modes.

voltage waveforms due to pure square wave or quasi-square wave approximations.

With such assumptions, the derived equivalent DAB circuit model, referred to the transformer's primary side, is presented in Fig. 3 (a). Here, the line inductor $L$ is the lumped inductance of the transformer leakages and any external series connected inductor, given by $L = L_p + L'_s \cdot (n_p^2/n_s^2)$.

For positive power transfer from the primary to the secondary side, depending on the relative values of the three control variables, $\varphi$, $\delta_p$ and $\delta_s$ (where, $\varphi, \delta_p, \delta_s \in \left[0, \frac{\pi}{2}\right]$), the converter operates in five distinct zones [11], each corresponding to different profiles of the inductor voltage $v_L$ (= $v_p - v_s$) and $i_L$. Using this conventional Ideal Lossless Model (ILM) [11], the average transferred power, $P_{o,ILM}$, for all operating modes can be determined as presented in Table I, where the output dc voltage gain is denoted as $m = V_{in}/V_{out}$.

Thus, it appears that with a known load power and voltage information, the lumped inductance $L$ can be easily estimated under any DAB operating mode, employing TPS, DPS, or SPS modulation. While few existing works [26-27] utilize this method for inductance estimation under SPS modulation, the consideration of ideal circuit elements and a lossless DAB model often leads to significant errors in the inductance calculation (ranging from 5% to 20%, depending on the operating conditions). Moreover, synthesizing the loss optimized DAB modulation based on such conservative models also leads to non-optimal converter operating points and loss of soft switching, resulting in less efficient power conversion. Therefore, an improved modeling approach that addresses the limitations of the conventional method is of utmost importance, as proposed below.

*B. Proposed Non-ideality Inclusive Frequency domain Model (NIFDM):*

In practical applications, the parasitic elements associated with the DAB power stage components particularly on-state switch resistances, ac resistances of the inductor and transformer windings, equivalent series resistance (ESR) of the capacitors, inter and intra-winding parasitic capacitances of the transformer play a significant role in shaping the HF brunch currents, and overall power flow in the system. Thus, in this proposed modeling method, the HF transformer is modeled using a T-type circuit with parasitic capacitances $C_{ip}$, $C_{is}$, and $C_{ps}$, denoting the primary intra-winding, secondary intra-winding, and primary-secondary inter-winding capacitances, respectively (Fig. 4(a)). Such capacitive effects become more prominent in the case of planar transformer configuration [32] where the overlap areas between any successive layers of turns are typically larger than a wire-wound transformer with Litz winding or hook-up wire-based winding. Further, to avoid the mode-dependent time-domain analysis of the DAB, a Fourier series-based frequency domain modeling approach is incorporated. For the $k$-th harmonic of the switching frequency, the complete parasitic component inclusive DAB equivalent circuit is presented in Fig. 4(a). The $k$-th harmonic of the primary and secondary bridge voltages are represented as,

$$\overrightarrow{V_{p,k}} = \frac{4V_{in}}{k\pi}\cos(k\delta_p) \angle 0^o \quad (1)$$

$$\overrightarrow{V_{s,k}} = \frac{4V_o}{k\pi}\cos(k\delta_s) \angle -k\varphi. \quad (2)$$

TABLE I
DAB POWER FLOW ANALYSIS UNDER DIFFERENT OPERATING MODES

| Operating Modes | Condition | Transferred Power, $P_{ILM}(p.u.)$; $P_{base} = V_{in}^2/2\pi f_{sw}L$ |
|---|---|---|
| Mode-1 | $\varphi \geq \delta_p + \delta_s;$ $\delta_p + \delta_s \leq \frac{\pi}{2}$ | $m\left[\varphi\left(1 - \frac{\varphi}{\pi}\right) - \frac{\delta_p^2 + \delta_s^2}{\pi}\right]$ |
| Mode-2 | $\frac{\pi}{2} \geq (\varphi_s - \varphi_p);$ $\varphi \geq \pi - (\delta_p + \delta_s)$ | $\frac{2m}{\pi}\left(\frac{\pi}{2} - \delta_p\right)\left(\frac{\pi}{2} - \delta_s\right)$ |
| Mode-3 | $\|\delta_p - \delta_s\| \leq \varphi;$ $\varphi < \min\begin{bmatrix}(\delta_p + \delta_s), \\ (\pi - \delta_p - \delta_s)\end{bmatrix}$ | $m\left[\varphi\left(1 - \frac{\varphi}{2\pi}\right) - \frac{\varphi(\delta_p + \delta_s)}{\pi} - \frac{(\delta_p - \delta_s)^2}{2\pi}\right]$ |
| Mode-4 | $\varphi < \|\delta_p - \delta_s\|;$ $\delta_s > \delta_p$ | $m\varphi\left(1 - \frac{2\delta_s}{\pi}\right)$ |
| Mode-5 | $\varphi < \|\delta_p - \delta_s\|;$ $\delta_p > \delta_s$ | $m\varphi\left(1 - \frac{2\delta_p}{\pi}\right)$ |

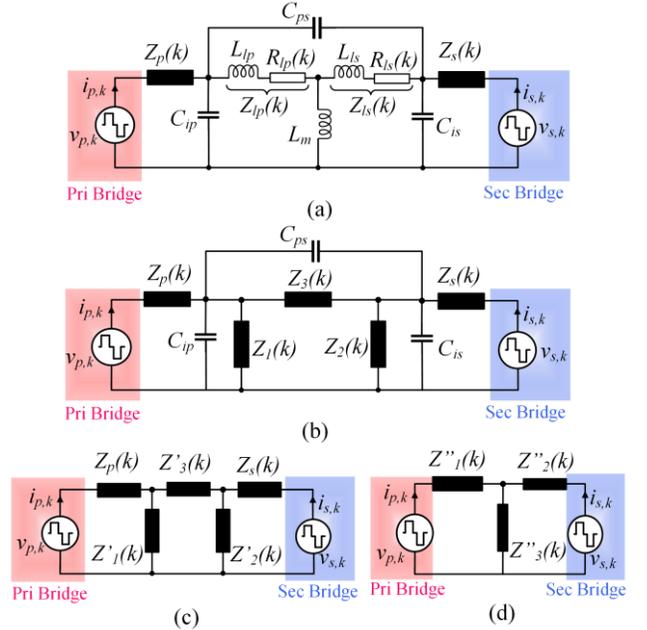

Fig. 4. Evolution of the parasitic inclusive passive element network of the DAB converter: (a) Complete DAB equivalent network including R-L-C parasitics; (b) star-delta converted version of the parent network in (a); (c) equivalent 2-port generalized network; (d) simplified T-network equivalent of DAB circuit.

In the equivalent circuit, $Z_p(k)$ and $Z_s(k)$ denotes the $k^{th}$ harmonic impedance that are series equivalents of tank inductances and resistive non-idealities on the primary and secondary sides, respectively. Thus, $Z_p(k) = R_p(k) + jk\omega_s L_p$ and $Z_s(k) = R_s(k) + jk\omega_s L_s$, ($\omega_s = 2\pi f_{sw}$, with $f_{sw}$ being the converter switching frequency) where, $L_p$ and $L_s$ are the primary and secondary side series inductances. $R_p(k)$ and $R_s(k)$ represent the primary and secondary side lumped equivalent resistances, consisting of the ac resistance of external inductor windings corresponding to $k$-th harmonic frequency, trace resistance, ESR of dc blocking capacitors, and switch on-state resistances. $L_m$, $L_{lp}$ and $L_{ls}$ represent the magnetizing inductance, primary and secondary leakage inductances of the transformer, respectively. Further, the ac resistances of the transformer's primary and secondary windings are denoted as $R_{lp}$ and $R_{ls}$. The frequency dependent or independent circuit elements present in Fig. 4(a) can be identified by individually characterizing the fabricated transformer, inductors, capacitors and switching devices. The network in Fig. 4(b) is a star-delta transformed circuit of the Y-type resistive-inductive network in Fig. 4(a), with transformed impedances:

$$\left. \begin{array}{l} Z_1(k) = Z_{lp}(k) + Z_{lm}(k) + \frac{Z_{lp}(k)Z_{lm}(k)}{Z_{ls}(k)} \\ Z_2(k) = Z_{ls}(k) + Z_{lm}(k) + \frac{Z_{ls}(k)Z_{lm}(k)}{Z_{lp}(k)} \\ Z_3(k) = Z_{lp}(k) + Z_{ls}(k) + \frac{Z_{lp}(k)Z_{ls}(k)}{Z_{lm}(k)} \end{array} \right\} . \quad (3)$$

Here, $Z_{lp}(k) = R_{lp}(k) + j\omega L_{lp}$; $Z_{ls}(k) = R_{ls}(k) + j\omega L_{ls}$; and $Z_{lm} = j\omega L_m$.

The shunt combinations of Z-C pairs such as $(Z_1, C_{ip})$, $(Z_2, C_{is})$, and $(Z_3, C_{ps})$ can be condensed to form the further simplified network shown in Fig. 4(c), which then after undergoing a delta-star transformation of $(Z_1', Z_2', Z_3')$ network produces the network in Fig. 4(d). The new impedances of the circuits in Fig. 4(c) and (d) are as follows:

$$\left. \begin{array}{l} Z_1'(k) = Z_1(k) || \frac{1}{jk\omega_s C_{ip}} \\ Z_2'(k) = Z_2(k) || \frac{1}{jk\omega_s C_{is}} \\ Z_3'(k) = Z_3(k) || \frac{1}{jk\omega_s C_{ps}} \end{array} \right\} \quad (4)$$

$$\left. \begin{array}{l} Z_1''(k) = Z_p(k) + \frac{Z_1'(k)Z_3'(k)}{Z_1'(k)+Z_2'(k)+Z_3'(k)} \\ Z_2''(k) = Z_s(k) + \frac{Z_2'(k)Z_3'(k)}{Z_1'(k)+Z_2'(k)+Z_3'(k)} \\ Z_3''(k) = \frac{Z_1'(k)Z_2'(k)}{Z_1'(k)+Z_2'(k)+Z_3'(k)} \end{array} \right\}. \quad (5)$$

Based on the final simplified DAB circuit model, given in Fig. 4(d), the $k$-th harmonic current vectors of the primary and secondary sides can be found using (6) and (7).

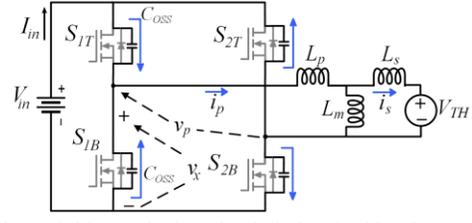

Fig. 5. Primary bridge equivalent circuit during deadtime interval between $S_{1B}$ and $S_{2T}$ turn-off and turn-on of $S_{1T}$ and $S_{2B}$.

$$\overrightarrow{I_{p,k}} = \frac{\overrightarrow{V_{p,k}}}{Z_1''(k)+Z_2''(k)||Z_3''(k)} - \frac{\overrightarrow{V_{s,k}}}{Z_1''(k)+Z_2''(k)+\frac{Z_1''(k)\cdot Z_2''(k)}{Z_3''(k)}} \quad (6)$$

$$\overrightarrow{I_{s,k}} = \frac{\overrightarrow{V_{s,k}}}{Z_2''(k)+Z_1''(k)||Z_3''(k)} - \frac{\overrightarrow{V_{p,k}}}{Z_1''(k)+Z_2''(k)+\frac{Z_1''(k)\cdot Z_2''(k)}{Z_3''(k)}} \quad (7)$$

The instantaneous currents of the primary and secondary bridges can be expressed using superposition of the harmonic components up to any order 'm', such as

$$i_x(t) = \sum_{k=1,3,5\ldots}^{2m+1} |\overrightarrow{I_{x,k}}| \sin(k\omega_s t + \angle \overrightarrow{I_{x,k}}). \quad (8)$$

The RMS value of the HF currents can be attained as

$$I_{x,RMS} = \sum_{k=1,3,5\ldots}^{2m+1} \frac{1}{2} |\overrightarrow{I_{x,k}}|^2. \quad (9)$$

Further, the active power sourced by the primary and secondary ac bridges are formulated employing (10) and (11).

$$P_{p,ac} = \sum_{k=1,3,5\ldots}^{2m+1} \frac{1}{2} |\overrightarrow{V_{p,k}}||\overrightarrow{I_{p,k}}| \cos(\angle \overrightarrow{I_{p,k}}) \quad (10)$$

$$P_{s,ac} = \sum_{k=1,3,5\ldots}^{2m+1} \frac{1}{2} |\overrightarrow{V_{s,k}}||\overrightarrow{I_{s,k}}| \cos(k\varphi + \angle \overrightarrow{I_{s,k}}) \quad (11)$$

The difference between $|P_{p,ac}|$ and $|P_{s,ac}|$ relates to the total conduction loss incurred by the resistive non-idealities in the converter.

Although the above method of non-ideality inclusive DAB modeling approach is significantly better compared to traditional ILM, the assumption of zero-deadtime based instantaneous switching transition of bridge voltages $v_p$ and $v_s$ leads to erroneous calculations of the HF current waveshapes and power flow. Thus, an effort is further made in this work to consider the effects of non-ideal switching voltage transitions during dead-time intervals. Here, the dead-time period of $T_d$ is assumed to occur at the rising edge of the gate pulses. In Fig. 5, an equivalent DAB circuit is displayed that corelates to a mode transition where the primary bridge voltage $v_p$ goes from $-V_{in}$ to $V_{in}$ during the deadtime duration between turn-off of $S_{1B}$ and $S_{2T}$ and turn-on of $S_{1T}$ and $S_{2B}$. During the deadtime $T_d$, the drain-source voltage of $S_{1T}$ and $S_{2B}$ is denoted by $(V_{in} - x)$, while the same for $S_{1B}$ and $S_{2T}$ is denoted by $x$ that varies from 0 to $V_{in}$. $V_{TH}$ represents the secondary bridge voltage $v_s$ during the switching instant reflected to primary side. The following equations can be formulated by applying KVL on the equivalent circuit.

$$v_p = 2x - V_{in} = L_p i_p' + L_m(i_p' - i_s') \quad (12)$$

$$L_m(i'_p - i'_s) = L_s i'_s + V_{TH} \quad (13)$$

Sum of the currents responsible for $S_{1T}$ body capacitor discharging and $S_{1B}$ body capacitor charging is $i_p$:

$$i_p = -2C_{oss}x'. \quad (14)$$

Using the set of ordinary differential equations given in (12)-(14), the following relation can be stated.

$$x'' + \frac{x}{C_{oss}L_{eq}} = \frac{V_{eq}}{C_{oss}L_{eq}} \quad (15)$$

Here, $L_{eq} = L_p + L_m || L_s$ and $V_{eq} = V_{in} - \frac{L_m}{L_m + L_s} V_{TH}$. The general solution of (15) can be expressed as

$$x = V_{eq} + A\sin(\omega_0 t) + B\cos(\omega_0 t), \quad (16)$$

where $\omega_0 = \frac{1}{\sqrt{L_{eq}V_{eq}}}$. Using the initial conditions of $x(0) = 0$, and $x'(0) = -\frac{i_p(0)}{2C_{oss}} = A\omega_0$, the unknown coefficients of (16) can be found as: $A = -\frac{i_p(0)}{2\omega_0 C_{oss}}$ and $B = -V_{eq}$. $i_p(0)$ can be calculated from (8) using the GHA model described earlier. Finally, switching voltage $x$ can be represented as,

$$x = V_{eq}[1 - \cos(\omega_0 t)] - \frac{i_p(0)}{\omega_0 C_{oss}} \cdot \sin(\omega_0 t). \quad (17)$$

Further, the equivalent circuit of Fig. 4 is only applicable for $i_p < 0$. Thus, keeping the direction of the inductor current in mind the solution of $v_p$ is represented in (18). Likewise, similar modeling can be performed for other switching commutation modes in a DAB and the expressions of the bridge voltages during deadtime interval can be derived. Such voltage waveform ($v_{p,dead,Sx}$) during deadtime before any MOSFET $S_x$ turns on, when added with the ideal quasi-square voltage waveform, yields the actual voltage waveform, as illustrated in Fig. 6. Using the relations given in (18), $v_{p,dead,Sx}$ can also be expanded in Fourier series with its k-th harmonic component to be $\overrightarrow{V_{p,dead,Sx,k}}$. Now, the modified k-th harmonic bridge voltages will be:

$$\overrightarrow{V_{p,mod,k}} = \overrightarrow{V_{p,ideal,k}} + \overrightarrow{V_{p,dead,S1T,k}} + \overrightarrow{V_{p,dead,S2T,k}} + \overrightarrow{V_{p,dead,S1B,k}} + \overrightarrow{V_{p,dead,S2B,k}}; \quad (19)$$

$$\overrightarrow{V_{s,mod,k}} = \overrightarrow{V_{s,ideal,k}} + \overrightarrow{V_{p,dead,S3T,k}} + \overrightarrow{V_{p,dead,S4T,k}} + \overrightarrow{V_{p,dead,S3B,k}} + \overrightarrow{V_{p,dead,S4B,k}}. \quad (20)$$

Here, $\overrightarrow{V_{p,ideal,k}} = \frac{4V_{in}}{k\pi}\cos(k\delta_p + 2k\omega_s T_d) \angle 0°$ and $\overrightarrow{V_{s,ideal,k}} = \frac{4V_o}{k\pi}\cos(k\delta_s + 2k\omega_s T_d) \angle -k\varphi$. Further, based on the modified bridge voltages the DAB inductor currents and power flows are recalculated using (6)-(11). The constructed HF DAB currents and bridge voltages are displayed in Fig. 7 showing a comparison between the ILM, the proposed NIFDM and the experimental results for a 160V-100V/200W power conversion. It is noticed that the NIFDM derived waveforms closely replicate the hardware waveforms for a partial and full ZVS scenario at the primary and secondary bridges, correspondingly. The error in estimating output power using the NIFDM at this operating condition is measured as 0.43% which is 16.2% lower compared to ILM estimated load power.

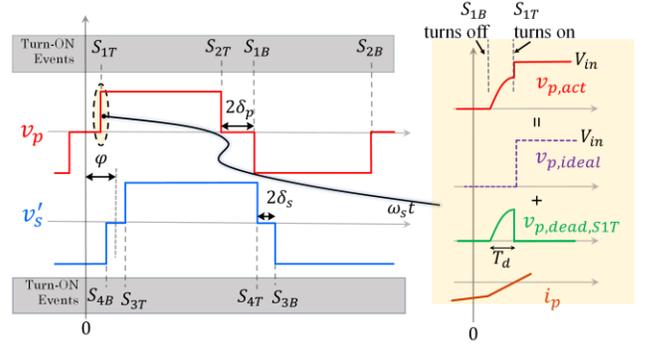

Fig. 6. Non-ideal switching transition waveform construction and definition of $v_{p,dead,S1T}$.

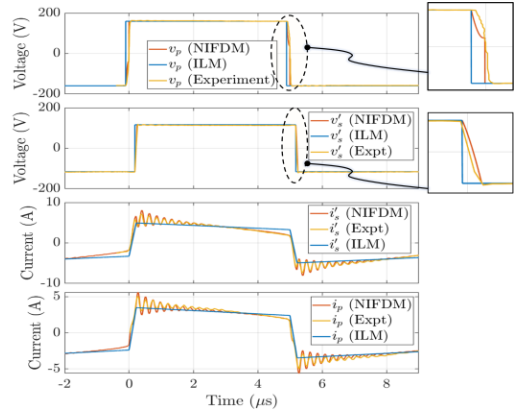

Fig. 7. Constructed DAB HF current and voltage waveshapes using NIFDM in comparison to experimental waveforms.

It is understandable from the previous analysis that for HF DAB converter, the circuit parasitics and deadtime effects can become critical in determining the power flow among the converter ports, the port current shapes; thus, it also influences the losses in the system. With the aim of synthesizing a fixed frequency TPS modulation scheme for DAB that targets minimizing the system losses, precise loss models of the individual circuit elements as function of the control parameters and design parameters need to be formulated.

*1. Conduction Losses:*

$$v_p(t) = \begin{cases} \max\left[V_{in}, \left\{2V_{eq}(1-\cos(\omega_0 t)) - \frac{i_p(0)}{\omega_0 C_{oss}} \cdot \sin(\omega_0 t) - V_{in}\right\}\right], & \text{if } i_p < 0; \ 0 < t < T_d \\ V_{in}, & \text{if } i_p > 0; 0 < t < T_d \\ V_{in}, & \text{if } t > T_d \end{cases} \quad (18)$$

With the power flow model established earlier, total conduction loss in the DAB power stage is derived as the function of the steady state control variables:
$$P_{cond}(\delta_p, \delta_s, \varphi) = |P_{p,ac}| - |P_{s,ac}|. \quad (21)$$

*2. Switching Losses:*

The switching loss in the MOSFETs can be computed by separately identifying the loss incurred by device's channel $v-i$ overlap during switching transition and the device's body capacitor, majorly $C_{oss}$ related loss when any incomplete-ZVS takes place.

The total switching loss in a DAB due to $v-i$ overlap is formulated in (22).

$$P_{sw,v-i}(\delta_p, \delta_s, \varphi) =$$
$$\sum_{i=p,s}\sum_{x=1}^{2}[2|v_{S_i,x}(\tau_{i,x,off})||i_i(\tau_{i,x,off})|f_{sw}t_{off,i,x} + 2|v_{S_i,x}(\tau_{i,x,on})||i_i(\tau_{i,x,on})|f_{sw}t_{on,i,x}]. \quad (22)$$

Here, $x$ denotes the leg undergoing switching event, it can be 1 or 2, based on the leading or lagging leg of any H-bridge. $\tau_{i,x,off}$ represents the time instant when the top device connected at leg-$x$ of bridge-$i$ is turned off. The corresponding drain-source voltage across that device and the drain current at $t = \tau_{i,x,off}$ are expressed as $|v_{S_i,x}(\tau_{i,x,off})|$ and $|i_i(\tau_{i,x,off})|$, respectively. Similarly, the variables that relate to the turn-on switching event of top device of leg- $x$ of bridge - $i$ are $|v_{S_i,x}(\tau_{i,x,on})|$, $|i_i(\tau_{i,x,on})|$ and $t_{on,i,x}$, representing the voltage across, current through device at turn-on instant and turn-on switching time. The $t_{off,i,x}$ and $t_{on,i,x}$ depends on device parameters ($C_{oss}$, $C_{iss}$, $C_{rss}$) and used gate resistances ($R_{G,on}$ and $R_{G,off}$) [33]. Moreover, the switching times and definition of $v_{S_i,x}$ for different DAB leg switches are highlighted in Table II.

The incomplete charge and discharge of the MOSFET's body capacitors ($C_{oss}$) also add on to major switching loss under partial or full hard-switching event. The calculation of the energy dissipation [34] during an incomplete ZVS-event at a switching leg of the DAB is portrayed below. Suppose $S_{1T}$ turns on while $C_{oss,S_{1T}}$ is still charged to $\Delta V$ voltage (see Fig. 8), which dissipates a certain amount of energy that can be derived by solving the energy balance of

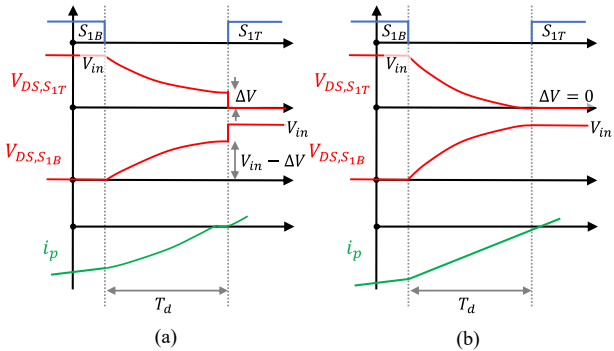

Fig. 8. Drain-Source voltage transition during (a) incomplete ZVS and (b) complete ZVS events.

TABLE II
DEFINITION OF VARIABLES USED IN (22) AND (23)

| Switch | $\tau_{i,x,on}$ | $\tau_{i,x,off}$ | $v_{S_i,x}(t)$ |
|---|---|---|---|
| $S_{1T}$ | $\delta_p + T_d$ | $\pi + \delta_p$ | $V_{in} - v_{p,mod}(t)$ |
| $S_{2T}$ | $\pi - \delta_p + T_d$ | $-\delta_p$ | $v_{p,mod}(t)$ |
| $S_{3T}$ | $\varphi + \delta_s + T_d$ | $\pi + \varphi + \delta_s$ | $V_{out} - v_{s,mod}(t)$ |
| $S_{4T}$ | $\varphi + \pi - \delta_s + T_d$ | $\varphi - \delta_s$ | $v_{s,mod}(t)$ |

$$E_{final} + E_{diss} = E_{delivered} + E_{initial}. \quad (23)$$

The residual voltage $\Delta V$ can be identified by solving the resonant circuit during the deadtime as outlined in (19) and (20). For any voltage $V_{DS}$ across any MOSFET, the stored energy at its nonlinear body capacitance $C_{oss}$ is $E_{oss}(V_{DS}) = \int_0^{V_{DS}} C_{oss}(v) \cdot v dv$ and can be calculated from the $C_{oss} - V_{DS}$ plot given in the device datasheet. Thus before $S_{1T}$ turns on, the total energy within the switching leg system due to the voltages across the $C_{oss}$ of $S_{1T}$ and $S_{1B}$, i.e., $\Delta V$ and $(V_{in} - \Delta V)$ is,

$$E_{initial} = E_{oss}(\Delta V) + E_{oss}(V_{in} - \Delta V). \quad (24)$$

After $S_{1T}$ turns on, $C_{oss}$ of $S_{1B}$ is charged to $V_{in}$, therefore, the final energy in the system is

$$E_{final} = E_{oss}(V_{in}). \quad (25)$$

Now, the amount of charge stored in $C_{oss}$ of any MOSFET due to $V_{DS}$ voltage present across it is denoted by $Q_{oss}(V_{DS})$ and can be calculated as, $Q_{oss}(V_{DS}) = \int_0^{V_{DS}} C_{oss}(v) \cdot dv$.

In order to charge the $C_{oss}$ of $S_{1B}$ from $(V_{in} - \Delta V)$ to $V_{in}$, the remaining charge is

$$\Delta Q_{S_{1B}} = Q_{oss}(V_{in}) - Q_{oss}(V_{in} - \Delta V). \quad (26)$$

This amount of charge has to be taken from the dc input source $V_{in}$. Thus,

$$E_{delivered} = \Delta Q_{S_{1B}} V_{in}. \quad (27)$$

Hence, the total dissipated energy due to the partial ZVS transition of $S_{1T}$ can be derived using (28) as,

$$E_{diss} = E_{oss}(\Delta V) + \Delta Q_{S_{1B}} V_{in} - [E_{oss}(V_{in}) - E_{oss}(V_{in} - \Delta V)]. \quad (28)$$

In case of complete soft-switching transition, $\Delta V$ will be 0, which yields $E_{diss} \to 0$. On the other hand, for complete hard-switching, $\Delta V \to V_{in}$; leading the total $C_{oss}$ related switching loss at that leg to $Q_{oss}(V_{in}) \cdot V_{in} \cdot f_{sw} = C_{Q,eq}(V_{in}) \cdot V_{in}^2 \cdot f_{sw}$. Here, $C_{Q,eq}(V_{in})$ is the charge-equivalent body-capacitance of the MOSFET at $V_{in}$ voltage.

Furthermore, extending this approach of switching loss calculation to all the DAB switching legs, a unified total $C_{oss}$ related switching loss expression is derived and presented in (29).

$$P_{SW,Coss}(\delta_p, \delta_s, \varphi) = \sum_{i=p,s}\sum_{x=1}^{2} 2\big[E_{oss,i}(|v_{S_i,x}(\tau_{i,on,x})|) + \{Q_{oss,i}(V_i) - Q_{oss,i}(V_i - |v_{S_i,x}(\tau_{i,on,x})|)\}\cdot V_i - \{E_{oss,i}(V_i) - E_{oss,i}(V_i - |v_{S_i,x}(\tau_{i,on,x})|)\}\big]\cdot f_{sw}. \quad (29)$$

Here, $E_{oss,i}(|v_{S_i,x}(\tau_{i,on,x})|)$ denotes the energy stored in the nonlinear capacitance $C_{oss}$ of $i$-th bridge MOSFET due to the presence of $|v_{S_i,x}(\tau_{i,on,x})|$ voltage across it during turn-on instant. The voltage $|v_{S_i,x}(\tau_{i,on,x})|$ can be derived from the established NIFDM model of the DAB using (19)-(20) and Table II. In (29), $V_i$ is the dc link voltage; thus, for the primary and secondary H-bridge switches, it will take a value of $V_{in}$ and $V_{out}$, respectively.

Now, the total switching loss in the DAB switching network can be identified as sum of (22) and (29):

$$P_{SW}(\delta_p, \delta_s, \varphi) = P_{SW,v-i}(\delta_p, \delta_s, \varphi) + P_{SW,Coss}(\delta_p, \delta_s, \varphi). \quad (30)$$

It is evident from the derived switching loss model that the DAB phase shift variables $(\delta_p, \delta_s, \varphi)$ play the crucial role in shaping the bridge voltages and the inductor current and thereby directly influence the achievement of ZVS and the losses incurred during partial or hard-switching events. Consequently, if $P_{SW}(\delta_p, \delta_s, \varphi)$ is targeted for minimization using an optimization program, with $\delta_p, \delta_s$ and $\varphi$ as the decision variables, the optimal control parameters will aim to produce an inductor current waveform that facilitates ZVS by ensuring that the residual drain-source voltage before switching, i.e., $\Delta V$ approaches zero for each of the four switching legs. However, in a fixed-$f_{sw}$ DAB design with a significantly high $L_m$, simultaneously achieving ZVS for all four switching legs across the entire power flow and gain range is mathematically infeasible, as noted in previous studies [15]-[16]. This limitation arises due to presence of finite non-linear $C_{oss}$ of the MOSFETs and their corresponding critical ZVS inductor current requirements. Due to this limitation, minimizing $P_{SW}(\delta_p, \delta_s, \varphi)$ does not guarantee all-ZVS operation across the full operational range of DAB. However, it maximizes the full-ZVS-operable range of the converter by optimizing the control variables to prioritize ZVS where possible. This approach maximizes the converter's efficiency and performance under a broader range of operating conditions while acknowledging the inherent trade-offs in the design.

### 3. Magnetic Core Losses:

With the calculated peak flux density $B_m$ through the series inductor and transformer cores and operating $f_{sw}$ of $100kHz$, the core losses in the utilized magnetic cores with 'R' material manufactured by Mag-Inc [35], are computed by the data fitted Steinmetz's equation function given in (31).

$$P_{core} = 0.00353 \cdot f_{sw}^{1.42} \cdot B_m^{2.88} \cdot T_c \cdot V_{core} \; mW \quad (31)$$

In (23), $f_{sw}$ is in Hz, $B_m$ is in T, $V_{core}$ is in $cm^3$ and $T_c$ is the temperature dependent term expressed as, $T_c = 1.97 - 0.02226 \cdot T + 0.000125 \cdot T^2$, where $T$ is temperature of the magnetic core in °C.

Following this, the total loss in the DAB converter power stage is computed as

$$P_{DAB\;loss}(\delta_p, \delta_s, \varphi) = P_{cond}(\delta_p, \delta_s, \varphi) + P_{SW}(\delta_p, \delta_s, \varphi) + P_{core}(\delta_p, \delta_s, \varphi). \quad (32)$$

Upon precise quantification of the DAB losses, the output side load power will be yielded after subtracting the total core loss and the switching loss corresponding to secondary side H-bridge $(P_{SW}(\delta_p, \delta_s, \varphi)|_s)$ from $P_{s,ac}$.

$$P_{s,out}(\delta_p, \delta_s, \varphi) = |P_{s,ac}| - P_{mag}(\delta_p, \delta_s, \varphi) - P_{SW}(\delta_p, \delta_s, \varphi)|_s. \quad (33)$$

Additionally, the total DAB input power will be derived by adding the primary H-bridge switching losses with $P_{p,ac}$.

$$P_{p,in}(\delta_p, \delta_s, \varphi) = P_{p,ac} + P_{SW}(\delta_p, \delta_s, \varphi)|_p. \quad (34)$$

Therefore, employing the proposed NIFDM, the losses in the DAB converter, input-output dc powers, HF current-voltage waveshapes can be precisely synthesized under any modulation strategy at any steady state operating point.

### III. SYNTHESIS OF CIRCUIT PARAMETER SENSITIVE LOSS OPTIMIZED DAB MODULATION

This section presents the formulation of a total converter loss-optimized fixed $f_{sw}$ TPS modulation scheme for DAB converter, leveraging the proposed NIFDM. This work formulates the optimal duty cycle $(\delta_p, \delta_s)$ parameters for the TPS controlled DAB converter as polynomial functions of the converter's key passive components, alongside its operating voltage and load conditions. This approach enables more adaptive and loss-optimal modulation across a wide range of operating conditions compared to the existing circuit parameter non-adaptive modulation techniques [11], [17], [19].

The DAB converter's power flow and overall system efficiency are heavily influenced by several critical circuit parameters, which can vary due to environmental changes or degradation over time such as the series inductances $L_p, L'_s$, the $R_{ds,on}$ of the primary and secondary side switches, $R_{ds,on,p}$, $R_{ds,on,s}$ and the ESR of the dc link or dc blocking capacitances, $R_{esr,p}$ and $R_{esr,s}$. Further, the lumped parasitic resistances impact the DAB power flow as can be understood from the established model given in Fig. 3(a). Thus, the primary and secondary side lumped resistances $R_{l,p}$ ($= R_{ds,on,p} + R_{esr,p}$) and $R_{l,s}(= R_{ds,on,s} + R_{esr,s})$ are considered here that are prone to vary. The main limitation is that the individual effect of any particular parasitic resistance cannot be captured from the power flow model. It is important to note that changes in the magnetizing inductance $L_m$ of the DAB transformer are not considered in this study, as its influence on power flow is negligible for a DAB unless $L_m$ becomes comparable to $L_p$, or $L_s$. While other circuit parameters, such as the dc-link capacitances, may also vary over time due to degradation or voltage stress, they do not affect the average power flow or the optimal phase-duty control parameters in a DAB. Therefore, this work focuses on adapting the loss-optimized modulation to account for variations in the series inductances and lumped resistances only.

Using the analytical framework provided by the proposed NIFDM, a numerical optimization routine is developed to minimize the total DAB losses under TPS modulation. The core of this process is illustrated in the flowchart (Fig. 9), and the optimization problem is formulated as follows:

A. *Decision Variables (x):* The DAB control parameters $\delta_p, \delta_s, \varphi$ construct the decision variables, which are constrained in the range of $[0, \frac{\pi}{2}]$.

B. *Cost Function, $f_{cost}(x)$:* In our study, the objective function to minimize is the total loss in the DAB power stage, denoted as $P_{DAB\ loss}(x)$, as given in (26). This encompasses conduction losses, switching losses, and core losses. Depending on the application requirements, $f_{cost}(x)$ can be also adapted to focus on any specific loss component, simplifying the optimization process. For instance, in low-voltage, high-current applications operating at switching frequencies in the tens of kHz or lower, conduction losses often dominate. In such scenarios, $f_{cost}$ can be equated to conduction loss or, $P_{cond}(\delta_p, \delta_s, \varphi)$ to streamline the optimization. Conversely, for high frequency (beyond tens of kHz) applications with high dc link voltage levels, achieving soft-switching and minimizing switching losses becomes critical. In these cases, $f_{cost}$ can be defined as switching loss or $P_{SW}(\delta_p, \delta_s, \varphi)$, as given in (29).

C. *Constraint:* The load demand at a given $V_{in}$ and $V'_{out}$, imposes the non-linear constrained on the optimization, presented as $P_{s,out}(x)$ ensuring that the output power meets the required load condition.

This optimization problem is solved to obtain the optimal TPS control variable set $x^*$ that minimize the total DAB loss $P_{DAB\ loss}(x^*)$ at a given operating point characterized by $V_{in}$, $V'_{out}$, and $P_{s,out}$. The optimization uses MATLAB's *fmincon* function for nonlinear constrained optimization. To avoid convergence to local minima, a multi-start approach is employed. This approach involves solving the optimization problem multiple times with different initial conditions, and the globally optimal solution is selected from the observed local minima values.

The optimization routine is iterated for equidistant values of $V'_{out}$, and $P_{s,out}$ over a wide range of DAB operating conditions, specifically $80V \leq V'_{out} \leq 150V$ (in steps of 5V) and $0 \leq P_{s,out} \leq 2kW$ (in steps of 50W), with a fixed $V_{in} = 160V$. Additionally, to account for variations in the circuit parameters, the optimization is repeated for different values of $L_p$, $L'_s$, $R_{l,p}$ and $R_{l,s}$. The parameter ranges and step sizes considered are as follows: $L_p \in [2\mu H, 6\mu H]$ in steps of $0.5\mu H$ (9 distinct values), $L'_s \in [1\mu H, 3\mu H]$ in steps of $0.5\mu H$ (5 distinct values), $R_{l,p} \in [20m\Omega, 80m\Omega]$ in steps of $10m\Omega$ (7 distinct values), and $R_{l,s} \in [5m\Omega, 30m\Omega]$ in steps of $5m\Omega$ (6 distinct values). As a result, a comprehensive dataset containing $852,390 = (11 \times 41 \times 9 \times 5 \times 7 \times 6)$ rows is generated, columns indexed by $L_p, L'_s, R_{l,p}, R_{l,s}, V'_{out}$, and $P_{s,out}$, along with the corresponding optimal TPS control variables $x^*$.

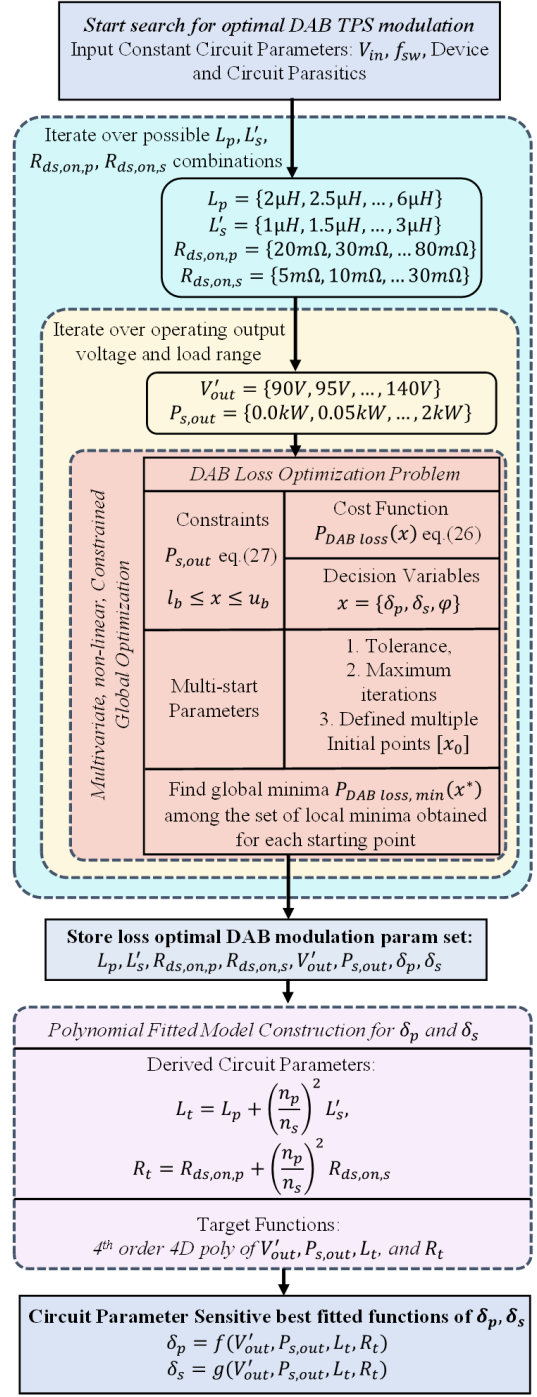

Fig. 9. Flowchart depicting the process of formulating polynomial models of loss optimized TPS control parameters as function of the DAB output voltage, load, series inductances and switching device on-resistances.

To implement these optimal TPS control variables in the controller hardware, the next step is to formulate polynomial functions for the duty cycle parameters as a function of the operating output voltage, output power, and the circuit parameters. It is important to mention that only the duty parameters are modeled here so that they can be implemented as feedforward terms in the DAB control system while the phase shift variable $\varphi$ is fetched from a proportional integral (PI) controller output that regulates the output voltage or current at the desired level [9]. To limit the complexity and the order of the polynomial functions, we introduce two derived variables that directly influence the optimal control variables $\boldsymbol{x}^*$. These variables are: primary referred total series inductance, $L_t = L_p + \left(\frac{n_p}{n_s}\right)^2 L'_s$ and lumped parasitic resistance, $R_t = R_{l,p} + \left(\frac{n_p}{n_s}\right)^2 R_{l,s}$. Given the high value of $L_m$, the lumped circuit parameters dictate the DAB power flow rather than the separate primary and secondary side circuit parameters. Following this, using the Penrose pseudoinverse method described in [36], the previously obtained dataset is fitted to generate fourth-order, four-dimensional polynomials for the optimal DAB duty parameters $\delta_p^*$ and $\delta_s^*$. These polynomials are expressed as functions of $V'_{out}, P_{s,out}, L_t$ and $R_t$: $\delta_p^* = f(V'_{out}, P_{s,out}, L_t, R_t)$ and $\delta_s^* = g(V'_{out}, P_{s,out}, L_t, R_t)$ such as $\delta_p^* = \sum_{0 \leq i+j+k+l \leq 4} a_{i,j,k,l} \cdot (P_{s,out})^i \cdot (V'_{out})^j \cdot (L_t)^k \cdot (R_t)^l$; and $\delta_s^* = \sum_{0 \leq i+j+k+l \leq 4} b_{i,j,k,l} \cdot (P_{s,out})^i \cdot (V'_{out})^j \cdot (L_t)^k \cdot (R_t)^l$. The selection of a 4th-order, 4-D polynomial fitting for the optimal duty cycle parameters balances the trade-off between the program execution time and modeling accuracy. A higher-order polynomial achieves greater modeling accuracy, reducing the efficiency compromise in the converter operation due to duty modeling error [36]. However, the computational burden of higher-order models can exceed the switching period (10 µs), leading to aliasing and control instability. For this work, a trade-off analysis shows that a 4th-order polynomial fitting results in a minimal efficiency compromise of 0.04%, compared to 0.16% and 0.39% for 3rd- and 2nd-order models, respectively. Benchmarking on a TMS320F28379D DSP confirms that the execution time for the 4th-order models is 7.7 µs, which is within the allowable 10 µs period.

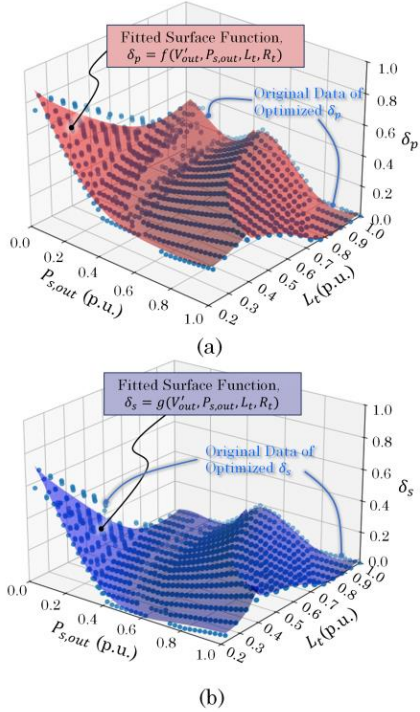

Fig. 10. Plots showcasing polynomial fitted models of converter loss optimal $\delta_p$ and $\delta_s$ and their original values with varying $L_t$ and $P_{s,out}$.

The variations between the fitted polynomials and the original data for the optimal duty variables, with respect to changes in $P_{s,out}$ and $L_t$, at a constant $V'_{out}$ of 100V and $R_t$ of 40mΩ, are shown in two 3D plots in Fig. 10. The high fitting accuracy of the nonlinear optimal functions $\delta_p$ and $\delta_s$ is evident from the plots, with calculated error variances of 3.1% and 2.7%, respectively. The plots also indicate that for a fixed gain and load, the optimal $\delta_p$ and $\delta_s$ varies significantly with changes in $L_t$, particularly in the light- to mid-load range. This demonstrates the importance of incorporating the value of $L_t$ into the calculation of modulation variables to ensure loss-optimized DAB operation.

Furthermore, to quantify and visualize the advantages of the proposed circuit parameter adaptive control over non-adaptive TPS control, the efficiency improvement $\Delta\eta$ % (computed based on NIFDM) using the proposed modulation is highlighted in Fig. 11 over a wide range of $L_t$ and load conditions. This plot assumes constant parameters of $V'_{out} = 100V$ and $R_t = 40m\Omega$. The results show that during light- to mid-load operation, the maximum efficiency improvement $\Delta\eta_{max}$ can reach up to 4.2%, which is primarily due to the loss of soft-switching and increased current stress observed with parameter non-adaptive TPS modulation. Also, $\Delta\eta$ increases with increase in $\Delta L_t$ due to increased difference in the optimal control variables compared to nominal $L_t$. A similar study has also been done to observe the $\Delta\eta$ using the proposed modulation with varied $R_t$, for a constant $L_t = 5\mu H$. With the observed $\Delta\eta_{max}$ of 0.6 %, it can be concluded that the impact of $R_t$ variation over modulation optimization is not as severe as $L_t$ change. However, it is beneficial to accommodate optimal duty models adaptable to both the circuit parameter $L_t$ and $R_t$ variations to ensure least loss converter operation.

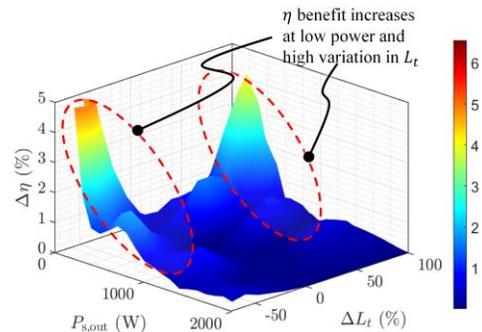

Fig. 11. Estimated efficiency improvement by applying proposed parameter adaptive TPS modulation, measured with respect to varied $L_t$ for wide operating load range.

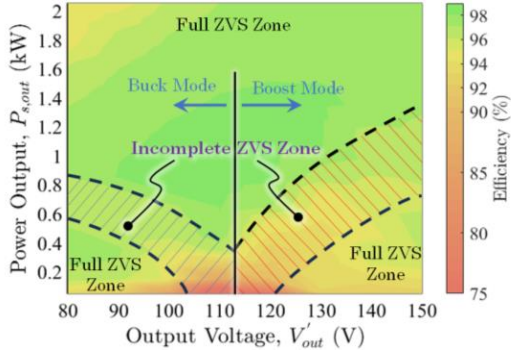

Fig. 12. Analytically calculated Soft-switching Region for the DAB ($L_t = L_{t,nom} = 11.024\ \mu H$) under proposed overall loss optimized TPS modulation. The incomplete ZVS range is marked as shaded region that appears due to presence of non-linear $C_{oss}$ of the primary and secondary bridge MOSFETs.

Therefore, the formulated polynomial representation of the TPS duty variables enables the implementation of the optimal TPS control in the hardware without the need for extensive lookup tables, ensuring real-time performance while accommodating changes in the DAB converter's operating conditions and circuit parameters.

Additionally, an observation on the complete soft-switching range of the DAB under the proposed TPS scheme is also made analytically and presented in Fig. 12. It can be noticed that due to fixed $f_{sw}$ control and the existence of finite $C_{oss}$ of the MOSFETs in both DAB bridges [15], the proposed modulation fails to attain complete ZVS across all the switches, when with increasing load the TPS transitions into DPS during buck and boost mode operation. At this incomplete ZVS region three or less switching legs experience partial ZVS while rest switching legs attain complete ZVS.

## IV. Physics Informed Neural Network for DAB Circuit Parameter Estimation

In Section III, we have established the necessity of accurately estimating the circuit parameters of the Dual Active Bridge (DAB) converter—specifically the lumped series inductance ($L_t$) and switch on-state resistance ($R_t$)—for effective implementation of parameter-adaptive optimal TPS modulation and for health monitoring purposes. To address this, we propose a Physics Informed Neural Network (PINN)-based parameter estimation algorithm. This approach allows real-time online estimation of $L_t$ and $R_t$ values for a physical DAB converter using the converter's sensed input/output voltages and currents ($V_{in}, V_{out}, I_{in}, I_{out}$), and steady-state control parameters ($\delta_p, \delta_s, \varphi$).

The PINN model is designed by combining two components: a data-driven neural network and a DAB physics model. The architecture of the PINN, shown in Fig. 13, is aimed at predicting the circuit parameters of the DAB converter by mapping the steady-state input/output variables to the parameters to be estimated. The DAB physics model, formulated based on the synthesized NIFDM, replicates the actual converter behavior with high accuracy. Incorporating this physics-based model accelerates the neural network's training process and reduces the required amount of training data.

### A. Training Data Collection and Pre-processing:

The PINN is designed to function efficiently with limited training data. The PINN achieves optimal performance in accurately predicting the circuit parameters when trained on a dataset entirely composed of hardware-collected data. For the hardware data collection, an automated test bench was developed, integrating DAB hardware components including the Texas Instruments TMS320F28379D control card, Preen ADG-L-160-75 DC power supply, Chroma 63212A-1200-480 DC electronic load, and Hioki PW3337 power meter. These components were controlled via a Python-based automation script on a PC. While this setup allows for the automatic collection of input/output voltage, current, and control variables, varying inductance values and replacing devices to obtain a wide variety of datasets is challenging due to manual intervention and a limited number of available devices with different $R_{ds,on}$ values. To improve the quality and diversity of the training data, additional datasets were generated using the NIFDM-based DAB model in MATLAB, where by simply varying the circuit parameters such as $L_p$, $L'_s$ and $R_{ds,on,p}$, $R_{ds,on,s}$, a richer dataset for training is generated. The NIFDM model has been priorly benchmarked to have 5.8% average modeling error compared to DAB hardware in terms of power flow calculation for any specific input/output voltage and phase-duty control operating point. For this study, we collected a total of 5000 data samples, comprising 1000 hardware datasets and 4000 simulated datasets from the NIFDM-based DAB model, which replicates real hardware performance with 4.6% average modeling error.

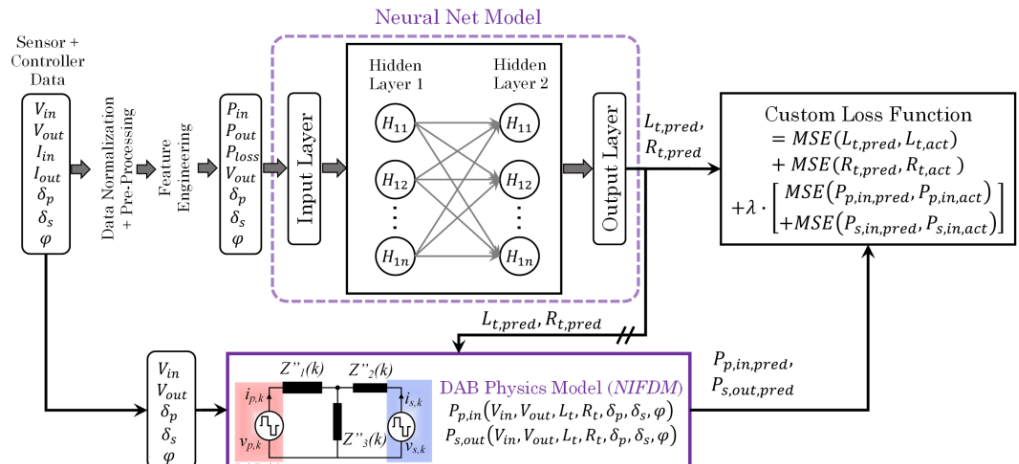

Fig. 13. Proposed Physics Loss Informed Neural Network Model used for DAB's parameter estimation.

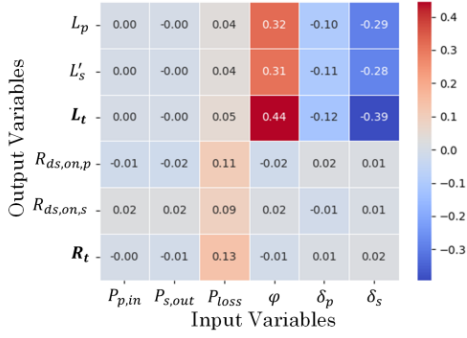

Fig. 14. Correlation Matrix of the Input and Output Variables within the training dataset.

Each collected dataset contains information on $L_p$, $L'_s$, $R_{ds,on,p}$, $R_{ds,on,s}$, $V_{in}$, $V_{out}$, $I_{in}$, $I_{out}$, $\delta_p$, $\delta_s$, and $\varphi$. Given that the magnetizing inductance $L_m$ is significantly larger (approximately 100 times) than $L_p$ and $L'_s$, very little current flows through $L_m$. Therefore, the lumped DAB inductor $L_t (= L_p + \left(\frac{n_p}{n_s}\right)^2 L'_s)$ and lumped on-state resistance $R_t$ ( = $R_{ds,on,p} + \left(\frac{n_p}{n_s}\right)^2 R_{ds,on,s}$ ) primarily govern the power flow. Hence, estimating $L_p$, $L'_s$, $R_{ds,on,p}$, and $R_{ds,on,s}$ individually is methodologically constrained; however, the lumped quantities $L_t$ and $R_t$ can be estimated as they exhibit clear relationships with the DAB input/output voltages, currents, and control variables.

Furthermore, since $R_t$ directly correlates with system losses, particularly conduction losses, it is expected to correlate strongly with the total power loss in the DAB power stage, $P_{loss}$ = $P_{p,in} - P_{s,out} = V_{in}I_{in} - V_{out}I_{out}$. Therefore, $P_{loss}$ is included as a derived feature from the training data set to improve the neural network's training. Ultimately, $L_t$ and $R_t$ represent the output variables of the neural network, while $V_{in}$, $V_{out}$, $I_{in}$, $I_{out}$, $P_{loss}$, $\delta_p$, $\delta_s$, and $\varphi$ are used as input variables.

The dataset is randomly split into training (70%), validation (20%), and test (10%) sets, ensuring homogeneity across the portions. The training data is used to learn the model, the validation data tested the model's generalization capabilities during training, and the test data provided performance evaluation. For consistent processing, all variables are normalized to the range [0, 1] using the formula:

$$Y_{nor} = \frac{Y - Y_{min}}{Y_{max} - Y_{min}} \quad (35)$$

where $Y$ represents the actual data, and $Y_{min}$ and $Y_{max}$ denote the minimum and maximum values of the variable, respectively. Fig. 14 illustrates the correlation matrix between the input and output variables for constant $V_{in}$ and $V_{out}$, generated using Pearson's correlation coefficient (ranging from -1 to 1). As anticipated, the lumped DAB inductance and resistance $L_t$ and $R_t$ are strongly correlated with $\varphi$ and $P_{loss}$, respectively.

TABLE III
ARCHITECTURE OF THE FINAL DATA-DRIVEN NN MODEL

| Model Size | Number of Neurons | Parameters | Value |
|---|---|---|---|
| Input Layer | 7 | Total Parameters | 15240 |
| Hidden Layer 1 | 64 | Trainable Parameters | 4994 |
| Hidden Layer 2 | 64 | Non-Trainable Parameters | 256 |
| Output layer | 2 | Optimizer Parameter | 9990 |

TABLE IV
HYPERPARAMETERS OF THE DATA-DRIVEN NN MODEL

| Parameter | Value | Parameter | Value |
|---|---|---|---|
| Loss Function | Mean Square Error (MSE) | Dropout Rate | 0.1 |
| Optimizer | Adam | Epochs | 230 |
| Learning Rate | 0.001 | Batch Size | 64 |
| L2 regularization | 0.01 | Metrics | Mean Absolute Error (MAE) |
| Activation Function | ReLU | Training Time | 13 secs |

### B. Neural Net Model:

Given the highly non-linear relationships between the input and output variables, a Multilayer Perceptron Neural Network (MLP-NN) was chosen for this work. In MLP-NNs, neurons are typically organized into fully connected layers. The network used in this study comprises an input layer with 7 neurons and an output layer with 2 neurons (providing $L_t$ and $R_t$). Between these layers, there are 2 hidden layers, each with 64 neurons. The complexity of the input and output relation to be estimated dictates the number of neurons and hidden layers. Each neuron processes a linear combination of inputs through an activation function. After a comparative analysis of activation functions such as sigmoid, hyperbolic tangent (tanh), and rectified linear unit (ReLU), ReLU was chosen for all layers except the output layer, which uses a linear activation.

The hyperparameters of the NN—such as learning rate, L2 regularization rate, batch size, and number of epochs—were optimized using a brute force grid search to minimize the estimation error. The final hyperparameters and the architecture of the trained network are summarized in Table III and Table IV, correspondingly.

### C. DAB Physics Model:

The DAB physics model is constructed based on the earlier developed non-ideality and loss inclusive DAB circuit model - NIFDM – that precisely replicates the hardware behavior at all operating conditions. In the proposed PINN approach, the NN - predicted $L_{t,pred}$ and $R_{t,pred}$ are used as input to the physics model. Using the inputs such as $V_{in}$, $V_{out}$, $\delta_p$, $\delta_s$, and $\varphi$ from the training data set and the NN-derived $L_{t,pred}$ and $R_{t,pred}$, the physics model estimates the DAB's total input and output power, $P_{p,in,pred}$ and $P_{s,out,pred}$, using (28) and (27), respectively.

*D. Custom Loss Function:*

The training of the proposed PINN uses a custom loss function, which linearly combines 'data loss' ($Loss_{data}$) and 'physics loss' ($Loss_{phys}$)' in a weighted sum, produced by the NN-model and DAB-physics model, respectively.

$$Loss_{custom} = [MSE(L_{t,pred}, L_{t,act}) + MSE(R_{t,pred}, R_{t,act})] \\ + \lambda \cdot \\ [MSE(P_{p,in,pred}, P_{p,in,act}) + MSE(P_{s,out,pred}, P_{s,out,act})] \quad (36)$$

Where $L_{t,pred}$ and $R_{t,pred}$ are the NN-predicted values of $L_t$ and $R_t$, respectively. $P_{p,in,pred}$ and $P_{s,out,pred}$ represent the DAB physics model – predicted input power and output power computed based on the $L_{t,pred}$ and $R_{t,pred}$ fetched from the NN output. The first term in (36) $[MSE(L_{t,pred}, L_{t,act}) + MSE(R_{t,pred}, R_{t,act})]$ highlights $Loss_{data}$, since it is generated by comparing the NN-predicted and actual $L_t$, $R_t$. The second term, $[MSE(P_{p,in,pred}, P_{p,in,act}) + MSE(P_{s,in,pred}, P_{s,in,act})]$ denotes the $Loss_{phys}$, which is formulated by comparing the physics model's predicted input and output power with the actual power values obtained from sensor data. The hyperparameter $\lambda$ balances the contributions of these two losses and is chosen using a grid search to minimize the average absolute error. The benefit of including $Loss_{phys}$ in the $Loss_{custom}$ is that minimizing $Loss_{phys}$ naturally helps minimizing $Loss_{data}$; thus, the NN gets trained faster and with less data availability. The hyperparameter $\lambda$ in the custom loss function determines the relative contributions of the data loss and the physics loss. Tuning $\lambda$ is critical to achieving a balance between data accuracy and physical consistency. The following guidelines were adopted for tuning $\lambda$:

1. Initial Scaling: The magnitudes of the data loss and physics loss were evaluated using $\lambda$. If either term dominated, $\lambda$ was adjusted to bring both terms to comparable scales.

2. Grid Search: A range of $\lambda$ values ($\lambda \in [0.01, 10]$) was tested, and the model's performance was evaluated on test data and physical constraints.

3. Dynamic Adjustment: During training, $\lambda$ was dynamically updated using the ratio of the data loss to the physics loss to maintain proportional contributions.

A grid search showed that $\lambda = 0.8$ achieved the equal contribution from the data loss and the physics loss in the presented case.

A very high value of $\lambda$ indicates more weightage on $Loss_{phys}$ thus, the NN may get trained faster while minimizing $Loss_{phys}$, at the cost of poor fitting of the data-driven model whose loss is neglected. Such NN when deployed standalone, will predict erroneous $L_t$ and $R_t$. On the other hand, if $\lambda$ value is kept very low, the $Loss_{phys}$ is neglected and the PINN becomes a data-driven NN that will take significant time and amount of data to train properly.

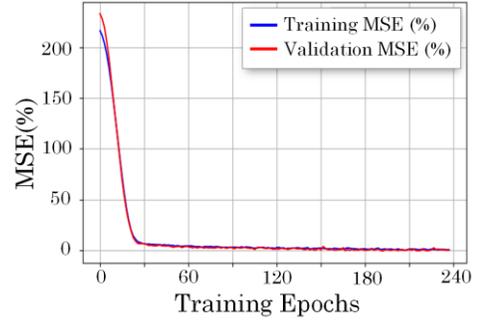

Fig. 15. Convergence Process of the PINN under training.

*E. Evaluation of PINN's Performance:*

The proposed PINN is developed on Keras TensorFlow 2.6 on an Intel i7-12900KF CPU with 32 GB of RAM. With this setup, the average training time of the NN comes near 30s. The backpropagation algorithm is utilized to update the NN weights and biases iteratively every time for 230 number of epochs. The training ends when the loss function, the MSE herein, stops decreasing. The validation dataset is used to detect over/underfitting. The initial weights of the network are chosen arbitrarily, and the entire training process (i.e., data splitting, training, and computation errors) is repeated multiple times to avoid local minimal solutions resulting from specific unfavorable initial conditions. The test-set is evaluated, and the obtained output is compared with the dataset output. If the performances of the ANN, in terms of error, are not sufficient, the ANN hyperparameters are updated iteratively.

Fig. 15 shows the convergence process of the final PINN that attains a minimum MSE of 1.23% and 1.44%, each corresponding to the training dataset and validation set, respectively after 230 epochs.

Further, the performance of the final trained NN is evaluated by comparing the estimated and actual values of $L_t$ and $R_t$, from the 'unseen' test data set. The results shown in

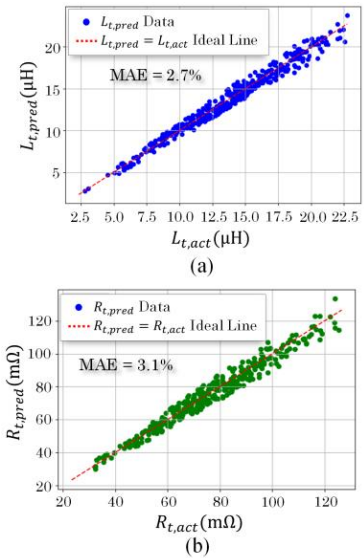

Fig. 16. Test data results evaluating the trained PINN. Plots highlight the estimated versus Actual (a) $L_t$ and (b) $R_t$.

Fig. 16 highlights that the mean absolute errors (MAEs) in $L_t$ and $R_t$ estimation are 2.7% and 3.1%, correspondingly. It is also noteworthy that, within the test data set, the actual values of $L_t$ spans over a wide range of possible inductances from 2.5 μH to 22.5 μH that corresponds to 0.2 times to 1.8 times of the nominal $L_t$. Thus, using such trained NN, a wide variation in $L_t$ and $R_t$ can be efficiently predicted. Furthermore, to justify the necessity of the physical model, we conducted a comparative analysis between the proposed PINN and a purely data-driven model. Key observations include: 1) when trained with 5000 datasets, the purely data-driven model achieved MAEs of 5.9% and 11% in predicting $L_t$ and $R_t$ on the unseen test dataset, which is 3.2% and 7.9% higher compared to the PINN's prediction errors, correspondingly. This demonstrates that the inclusion of the physical model improves modeling accuracy when a limited dataset is available. 2) The purely data-driven model required at least 20,000 data points to reach comparable accuracy, highlighting the PINN's data efficiency.

Another important point to note is that the proposed PINN structure is scalable and can be extended to estimate additional circuit parameters of the DAB, such as DC-link capacitances, magnetizing inductance ($L_m$), and others, in addition to $L_t$ and $R_t$. In such scenarios, this would require a larger and more diverse dataset, including detailed information on parameters like dc-link voltage ripple, to ensure accurate estimations. The architecture of the data-driven component of the PINN would need to be modified to achieve minimal error performance. Specifically, the number of nodes in the output layer would need to be increased to match the number of parameters to be estimated. This would result in a more complex network but would still retain the scalability and adaptability of the PINN framework.

## V. IMPLEMENTATION OF NN-IN-LOOP LOSS OPTIMIZED DAB MODULATION

This section presents the real-time implementation of the NN-based circuit parameter estimator integrated with the loss-optimized modulator for the DAB converter. The NN dynamically estimates key circuit parameters—lumped series inductance $L_t$ and the equivalent parasitic resistance $R_t$—which are used to adjust the TPS modulation variables for ensuring optimal performance under varying load conditions and circuit parameters. The control framework implementation is shown in Fig. 14 and is discussed in detail below.

*A. Neural Network Deployment:* The trained NN is implemented within a Python environment using TensorFlow/Keras on an Intel i7 PC. Although low-cost embedded platforms like STM32 microcontrollers, which offer native NN deployment tools (e.g., STM32Cube.AI), could be used, the Python implementation is chosen here due to ease of deployment and available hardware. The pretrained NN is saved as an .h5 file and is called during real-time operation. The NN takes as input the sampled values of operating $V_{in}, V'_{out}, I_{in}, I'_{out}, \delta_p, \delta_s, \varphi$ and outputs estimates of $L_t$ and $R_t$.

*B. DAB Modulator Implementation:* The loss-optimized DAB modulator is implemented directly on the Texas Instruments TMS320F28379D digital signal processor (DSP), which drives the DAB converter. Based on ADC-sampled sensor data of converter's input-output voltages, currents and the NN-estimated circuit parameters the modulator computes the optimal duty variables $\delta_p$ and $\delta_s$ using their polynomial fitted expressions. Each of these control variable computation takes approximately 8 μs on the DSP's 100 MHz core. To regulate the output voltage, a PI controller is employed that outputs the phase-shift variable $\varphi$, based on the error between the measured and target output voltage. The computed variables $\delta_p, \delta_s$ and $\varphi$ are then used to generate the gate pulses of H-bridge switching devices. The complete control loop, including the PI controller and polynomial function computation, operates within a total execution time of 18μs, allowing the converter to operate at 100 kHz with a down-sampling control cycle.

*C. Communication and Integration with NN:* To facilitate real-time updates of $L_t$ and $R_t$ between the Python-based NN and the DSP, a Serial Peripheral Interface (SPI) communication link is established using an Adafruit FT232H USB-to-SPI bridge. The FT232H device acts as the SPI master, while the TMS320F28379D DSP operates as the SPI slave. Given that $L_t$ and $R_t$ vary slowly over time, parameter estimation task is executed at intervals of approximately 30s. The data exchange process between the DSP and the PC is summarized as follows:

1. *Sensor Data Collection and Processing:* The DSP samples $V_{in}, V'_{out}, I_{in}, I'_{out}$ and processes these values including the control variables $\delta_p, \delta_s$ and $\varphi$ through digital low-pass filters of 10 kHz bandwidth to reduce switching frequency noise in these signals.

2. *Data Encoding for SPI Communication*: The processed sensor and control variables are scaled and converted into 16-bit fixed-point integers. Since SPI communication transmits 8 bits at a time, each variable is split into two 8-bit segments (high byte = first 8 bits and low byte = last 8 bits). The SPI communication operates at a modest baud rate of 10kHz, which minimizes noise coupling during data transmission and ensures reliable communication.

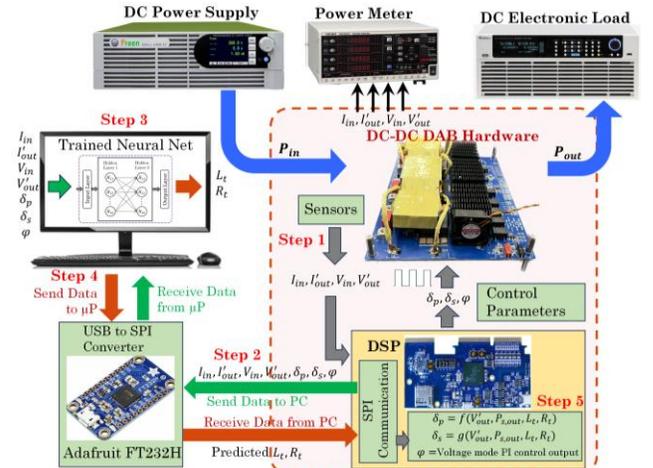

Fig. 17. NN-in-loop implementation of circuit parameter adaptive optimized DAB Modulation.

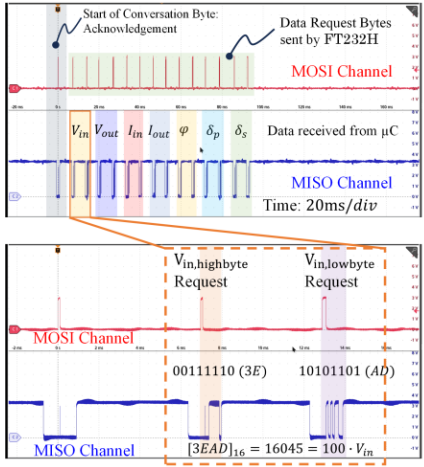

Fig. 18. NN-in-loop implementation of circuit parameter adaptive optimized DAB Modulation.

3. *Data Transmission to PC*: The SPI master initiates communication, prompting the DSP to transmit the requested data byte by byte. The high and low bytes for each 16-bit variable are transmitted sequentially. Once the data are received, the PC reconstructs the 16-bit integers, which are then passed to the NN algorithm.

Fig. 18 illustrates the SPI communication process during converter operation, showing how each of the 7 variables is transmitted byte by byte within a span of 95ms, which includes the data transfer time for 15 bytes with each byte transfer taking of 0.8ms (= 8 bits×1/10kHz) and the rest time of 5.93ms between two consecutive byte transfer. For example, it can be noticed that, while transmitting the input voltage $V_{in}$ data, the high byte ('3E') and low byte ('AD') are merged to form the 16-bit integer '3EAD' at the master end, which is equivalent to 16045 in decimal. After dividing by a scaling factor of 100, this yields the real-time $V_{in}$ value of 160.45V.

4. *NN Execution and Parameter Update*: Once the data streams have been received at the PC end, the NN processes the inputs and computes updated estimates for $L_t$ and $R_t$. These values are subsequently transmitted back to the DSP via the SPI bus in similar fashion, where they are used to modify the optimal modulation parameters $\delta_p$ and $\delta_s$ using multivariate polynomial regression functions that are already coded within the DSP, ensuring continuous loss-optimized operation.

This implementation ensures that the DAB converter maintains optimal efficiency, even under varying load conditions and changes in circuit parameters. The seamless communication between the PC and DSP enables real-time parameter updates, further enhancing the system's robustness and reliability.

## VI. EXPERIMENTAL VALIDATION AND BENCHMARKING

Upon finalizing the DAB control scheme, its performance is validated by implementing it in a laboratory designed and

TABLE V
DESIGN SPECIFICATION OF THE FABRICATED DAB CONVERTER

| Circuit Parameters | Specifications |
|---|---|
| Operating Voltages | Input: $V_1$ = 160 Vdc<br>Output: $V'_{out}$ = 80-150 Vdc; 115 Vdc nominal |
| Max Operating Power | $P_{s,out,max}$ = 2 kW |
| Prim Bridge Switches $S_{1T} - S_{2B}$ | GS66516T, 650V/60A/25mΩ @ 25°C GaN E-HEMT, 2 device in parallel per switch position, Driver: Si8271 |
| Sec Bridge Switches $S_{21} - S_{24}$ | EPC2215, 200V/42A/8mΩ @ 25°C GaN E-HEMT, 2 device in parallel per switch position, Driver: NCP51820AMNTWG |
| Primary DC Link Capacitor ($C_1$) | 40μF/250V<br>1 X B58033I5206M001<br>9 X C2220X225K251T<br>12 X CGA4J3X7R2E223K125AA |
| Secondary DC Link Capacitor ($C_2$) | 35μF/250V<br>10 X C2220X225K251T<br>4 X KTJ251B335M76BFT00<br>24 X CGA4J3X7R2E223K125AA |
| DC Blocking Capacitors ($C_{b1}, C_{b2}$) | 10μF/250V, 12μF/250V |
| Transformer | Turns Ratio: $n_1:n_2$ = 7:5<br>Magnetizing Inductance: $L_m$ = 770μH<br>Leakage Inductances of Prim and Sec: 0.24μH, 0.123μH<br>Core: 2X FR43808EC, R material, no airgap |
| Prim Inductor | Inductance: 2.15μH, 5.35μH<br>Winding: 4 turns, AWG 14 Litz wire<br>Core: 2 X 0R43618EC, R material |
| Sec Inductor | Inductance: 2.65μH<br>Winding: 4 turns, AWG 10 Litz wire<br>Core: 1 X 0R43618EC, R material |
| Switching frequency ($f_{sw}$) | 100 kHz |

fabricated proof-of-concept 2 kW DAB hardware, as shown in Fig. 19. The design targeted a wide gain range operation as specified in Table V which also lists the salient parameters of the converter. The port voltage specifications ($V_{in} = 160V$; $V'_{out} = 80 - 150V$) comply with a targeted isolated dc-dc power supply used in the international space station [37]. To enhance the power density and take advantage of high-frequency switching, all capacitors in the circuit were implemented using ceramic technology and the magnetic components were made using planar cores and PCB integrated windings. In order to reduce the losses, all the switches in the DAB H-bridges are realized using parallelly connected GaN

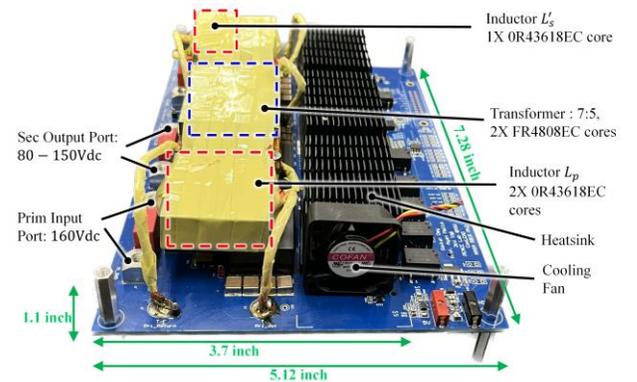

Fig. 19. Complete DAB converter prototype with marked magnetic components and thermal management system.

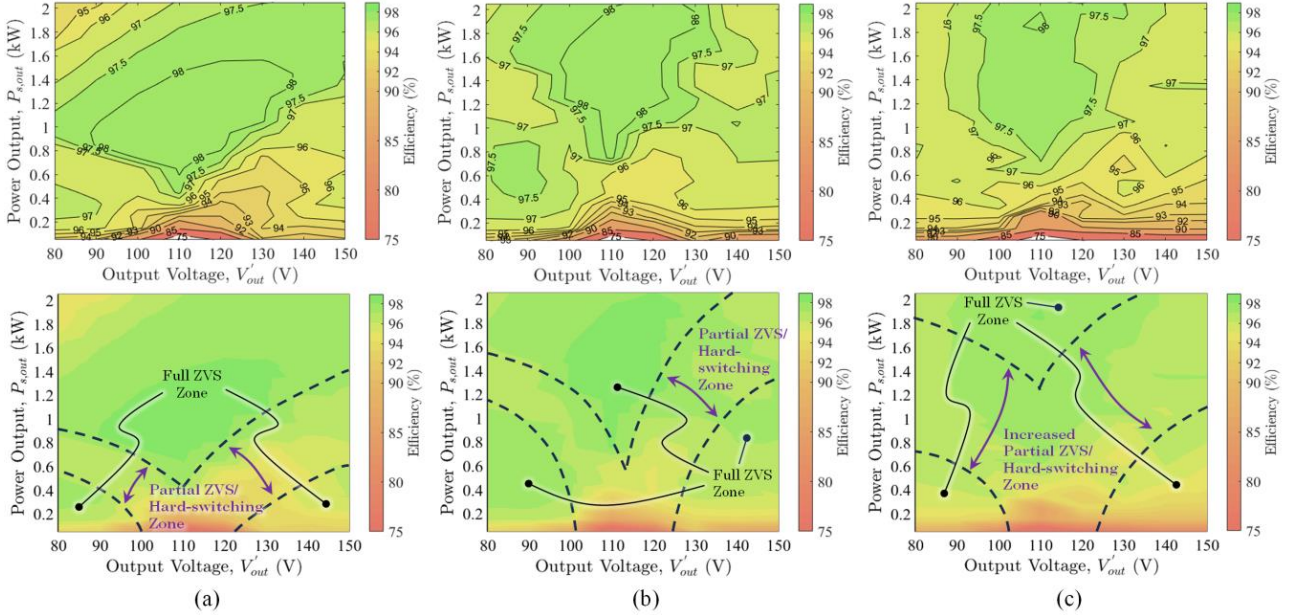

Fig. 20. Variations in DAB Efficiency with change in output power and load voltage when (a) $L_t = L_{t,nom}$; (b) $L_t = 0.7L_{t,nom}$ and $\delta_p = f(V'_{out}, P_{s,out}, L_t, R_t)$, $\delta_s = g(V'_{out}, P_{s,out}, L_t, R_t)$; and (c) $L_t = 0.7L_{t,nom}$ and $\delta_p = f(V'_{out}, P_{s,out}, L_{t,nom}, R_t)$, $\delta_s = g(V'_{out}, P_{s,out}, L_{t,nom}, R_t)$.

MOSFETs. The final construction details of all magnetic components: three-winding TAB transformer and the HF inductors are also summarized in Table V. The proposed DAB control strategy is implemented using a dual-core TMS320F28379D control card, connected at the bottom side of the depicted hardware. In order to check the converter's performance while varying $L_p$, the two series connected inductors $L_{p,1} = 2.15\mu H$ and $L_{p,2} = 3.2\mu H$ are implemented forming the total primary inductance $L_{p,nom} = 5.35\mu H$. A high current carrying MOSFET is connected in parallel to $L_{p,2}$, so that it can be shorted as and when need to emulate 60% reduction in $L_p$, or 30% reduction in total lumped series inductance $L_t$.

To experimentally benchmark the purpose and acquired performance benefit of introducing the parameter adaptive DAB modulation, the converter's efficiency and soft-switching range is monitored for full range variation in $V'_{out}$ and $P_{s,out}$. Such data is collected and presented in Fig. 20 for DAB operation under three different scenarios. Fig. 20(a): base case TPS operation with nominal circuit parameters; $L_t = L_{t,nom} = 11.024~\mu H$, $\delta_p = f(V'_{out}, P_{s,out}, L_t, R_t)$, $\delta_s = g(V'_{out}, P_{s,out}, L_t, R_t)$; Fig. 20 (b): NN-in-loop parameter adaptive control with reduced inductance, $L_t = 0.7L_{t,nom}$, $\delta_p = f(V'_{out}, P_{s,out}, L_t, R_t)$, $\delta_s = g(V'_{out}, P_{s,out}, L_t, R_t)$; and Fig. 20 (c): standard TPS operation with no parameter feedback with $L_t = 0.7L_{t,nom}$ and $\delta_p = f(V'_{out}, P_{s,out}, L_{t,nom}, R_t)$, $\delta_s = g(V'_{out}, P_{s,out}, L_{t,nom}, R_t)$. The efficiency data under such conditions is presented in form of color mapped contour plots in Fig. 20 (a), (b) and (c), respectively. The observed power flow zones that experience full-ZVS and partial-ZVS or hard-switching are also indicated in the corresponding plots. The summary of the observations are: 1) DAB achieves its highest efficiency of 98.2% at near unity voltage gain condition ($V'_{out} = 115V$) and at heavier load. As the voltage gain distances from unity, the efficiency decreases due to increased conduction and switching turn-off loss with non-flat current waveforms. 2) With implementation of loss optimized TPS, there appears three distinct zones of soft-switching: one at heavier load side more trapezoidal current shape is observed, two at lower load and voltage buck (voltage gain, $m < 1$) and boost ($m > 1$) side where TPS ensures ZVS with triangular current shape. In between these soft-switching zones, there appears a power flow band where full-ZVS cannot be attained due to insufficient inductive charge to discharge the switching leg $C_{oss}$. The width of this band depends on the stored charge in $C_{oss}$. Thus, at buck region, its width becomes narrower while in boost region, it becomes wider. 3) With reduced $L_t$, in Fig. 20(b), such non-ZVS zone shifts to higher power level due to reduced base power ( $P_{base} = V_{in}^2/2\pi f_{sw}L_t$ ) . But the ZVS zone characteristics remain the same due to parameter adaptive TPS implementation. 4) Since, Fig. 20(c) observes a TPS modulation with non-updated $L_t$, making the engaged $\delta_p$ and $\delta_s$ to be sub-optimal. This leads to non-optimal inductor current waveforms, resulting in loss of ZVS and increased RMS current. The increased zone of partial-ZVS/hard-switching is also evident from the plot. Comparing Fig. 20(b) and (c), the average and highest efficiency gain incurred by the engagement of parameter adaptive TPS is benchmarked to be 1.4% and 3.2%. This gain will become more prominent for high-voltage converters where switching loss becomes very dominant at lighter load levels, such as EV chargers.

Furthermore, the DAB performance with NN-in-loop optimized control with varying $L_t$ is validated for two different output voltage and load conditions:

*1)* $V_1 = 160V$, $V'_{out} = 100V$, and $P_{s,out} = 1.4kW$: Under this power conversion mode, the $L_p$ is purposefully reduced from $5.35\mu H$ to $2.15\mu H$, marking 30% reduction in $L_t$, while the converter waveforms are monitored over time. Fig. 21. (a) marks the DAB switching voltage and inductor current waveforms before the $L_t$ change. The implemented TPS algorithm ensures soft-switching at all four switching legs, with steady state control parameters being: $\varphi = 0.58$, $\delta_p = 0.285$ and $\delta_s = 0.1707$, with a measured power conversion efficiency of 97.80%. Immediately after the reduction of $L_t$ to $0.7L_{t,nom}$, the controller does not update the duty variables due to unavailability of the updated $L_t$ information from the NN that updates the estimates every 30s. However, the controller tracks the output voltage and load by varying $\varphi$ to 0.395. In the waveform shown in Fig. 21 (b), it can be observed that due to non-updated duty variables, soft-switching is missed at one of each H-bridge legs, resulting in an efficiency of 97.55%. Further, the circuit waveforms after incorporating the updated NN-estimated $L_t$ ($= 7.97\mu H$) and $R_t(= 63m\Omega)$ within the modulation variable calculation is presented in Fig. 21 (c). It can be observed that the soft-switching is regained alongside 8.5% reduction in inductor current RMS, leading to an improved system efficiency of 98.07%.

*2)* $V_1 = 160V$, $V'_{out} = 140V$, and $P_{s,out} = 0.7kW$: Under this power conversion mode, same change in $L_t$ is emulated, while observing the converter waveforms over time. Fig. 22. (a) marks the DAB switching voltage and inductor current waveforms before the $L_t$ change. The implemented loss optimized TPS algorithm cannot ensure soft-switching at all four switching legs, with steady state control parameters being: $\varphi = 0.23$, $\delta_p = 0.01$ and $\delta_s = 0.34$ (all in radian), with a reported power conversion efficiency of 95.70%. Immediately after the reduction of $L_t$ to $0.7L_{t,nom}$, with unchanged $\delta_p$ and $\delta_s$, and updated $\varphi$ of 0.19 (Fig. 22 (b)), soft-switching is again missed at both the primary side legs. This results in a measured DAB efficiency of 96.1%. Further, the circuit waveforms after incorporating the updated NN-estimated $L_t$ ($= 7.84\mu H$) and $R_t(= 53m\Omega)$ within the modulation variable calculation is presented in Fig. 22 (c). It can be observed that the soft-switching is reattained alongside a 5% reduction in inductor current RMS, leading to an increased system efficiency of 96.87%. This study establishes successful implementation of the NN-in-loop loss optimized DAB control.

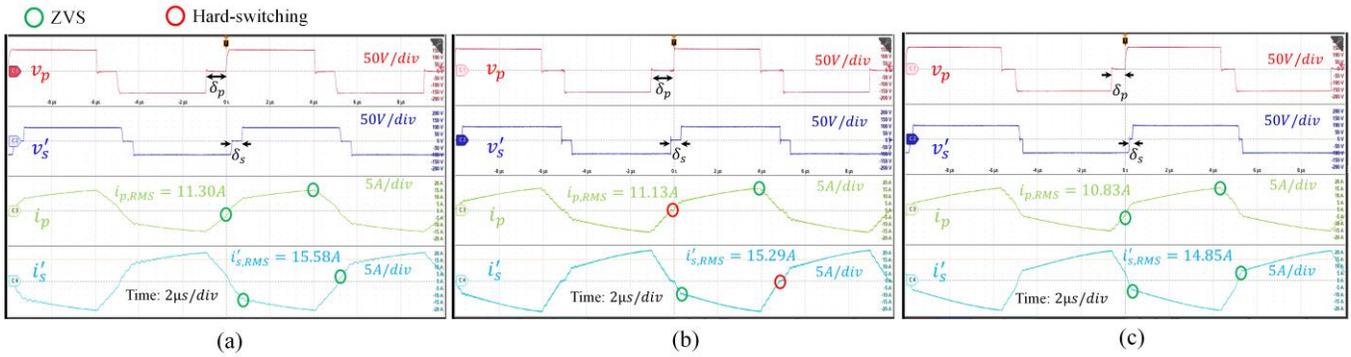

Fig. 21. Experimental steady state DAB converter waveforms for bridge voltages and currents at buck mode with $V_1 = 160V$, $V'_{out} = 100V$, and $P_{s,out} = 1.4kW$, demonstrating the variation of the primary side inductor $L_p$ during operation. (a) shows the waveforms with $L_p = 5.35\mu H$, (corresponding to $L_t = 11.024\mu H$), where the optimal TPS duty cycles calculated by the DSP are $\delta_p = 0.285$ and $\delta_s = 0.1707$. (b) depicts the waveforms right after $L_p$ is reduced to $2.15\mu H$, while the TPS duties remain unchanged due to the absence of updated $L_{t,pred}$ information. (c) displays the waveforms with updated TPS duties, $\delta_p = 0.241$ and $\delta_s = 0.067$, after the DSP incorporates the new $L_{t,pred}$ value from the NN loop.

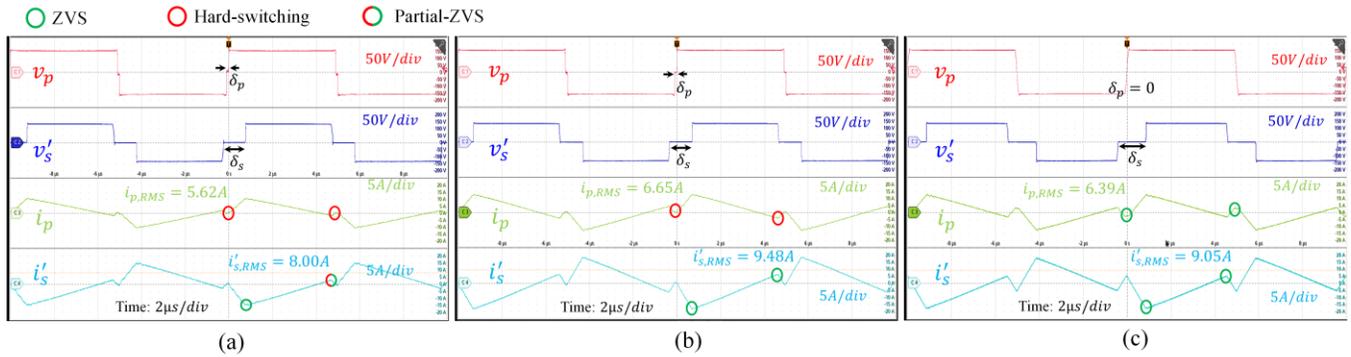

Fig. 22. Experimental steady state DAB converter waveforms for bridge voltages and currents at boost mode with $V_1 = 160V$, $V'_{out} = 140V$, and $P_{s,out} = 700W$, demonstrating the variation of the primary side inductor $L_p$ during operation. (a) shows the waveforms with $L_p = 5.35\mu H$, (corresponding to $L_t = 11.024\mu H$), where the optimal TPS duty cycles calculated by the DSP are $\delta_p = 0.01$ and $\delta_s = 0.34$. (b) depicts the waveforms right after $L_p$ is reduced to $2.15\mu H$, while the TPS duties remain unchanged due to the absence of updated $L_{t,pred}$ information. (c) displays the waveforms with updated TPS duties, $\delta_p = 0$ and $\delta_s = 0.42$, after the DSP incorporates the new $L_{t,pred}$ value from the NN loop.

Further, the efficiency of the DAB converter has been benchmarked for bidirectional power flow operation at various secondary side dc link voltages ($V_s'$) of 120V, 100V and 140V, while maintaining a fixed primary side dc link voltage of 160V. Fig. 23 illustrates the measured efficiency across a wide load range from -2kW to 2kW, under different TPS modulation schemes with varied circuit parameters. The results clearly demonstrate that the proposed parameter-adaptive TPS scheme achieves a peak efficiency improvement of 4% and an average efficiency gain of 1.8% at these voltage levels, enabled by optimized winding current RMS values and an expanded ZVS range.

Real-time performance validation of the NN-in-loop DAB modulation is further conducted by benchmarking the real-time communicated data received by the NN and its estimated $L_t$ and $R_t$ parameters while the hardware operated under controlled and varied load conditions. These tests were performed with different circuit inductances and device resistances (see Fig. 24). The device resistance was varied by using two and one device in parallel on the primary and secondary sides, respectively. Additionally, a low-value resistance was placed in series with the DAB inductor to emulate various resistance setups. The results demonstrate that, under varied operating conditions, the NN successfully predicted $L_t$ and $R_t$ with maximum error of 4.3% and 9.2%, respectively, while the mean errors are only 0.87% and 1.4%, correspondingly.

Further, to compare the estimation error of $L_t$ using the proposed ANN technique with the existing analytical method based on the time-domain DAB model [11], the predicted $L_t$ values from both approaches are plotted in Fig. 24(j). It is observed that the analytically estimated $L_t$ exhibits an average absolute error of 10%, primarily due to the exclusion of circuit non-idealities in the modeling process. These non-idealities include device on-resistance, dead-time effects, transformer parasitics, and losses in power stage components such as switching devices and passive elements. In contrast, the NN-based parameter estimation process demonstrates substantially higher accuracy due to its ability to account for these non-idealities and mimic the behavior of real hardware. This enhanced accuracy makes the NN approach highly suitable for implementing parameter-adaptive modulation schemes in DAB converters.

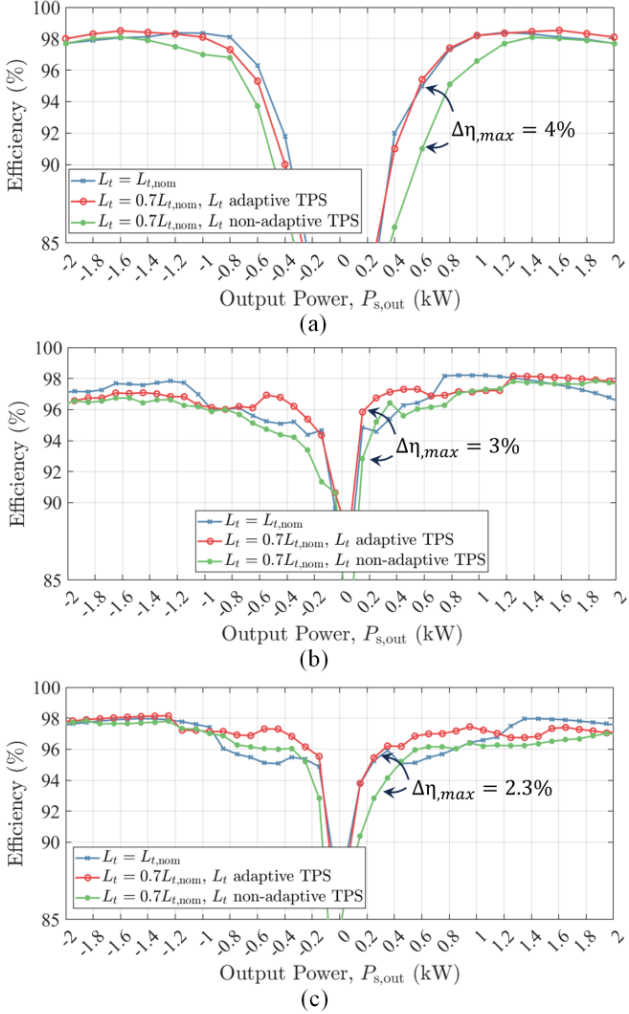

Fig. 23. DAB efficiency vs. output power under different modulation strategies at different output voltages: (a) $V_s' = 120V$, (b) $V_s' = 100V$, and (c) $V_s' = 140V$.

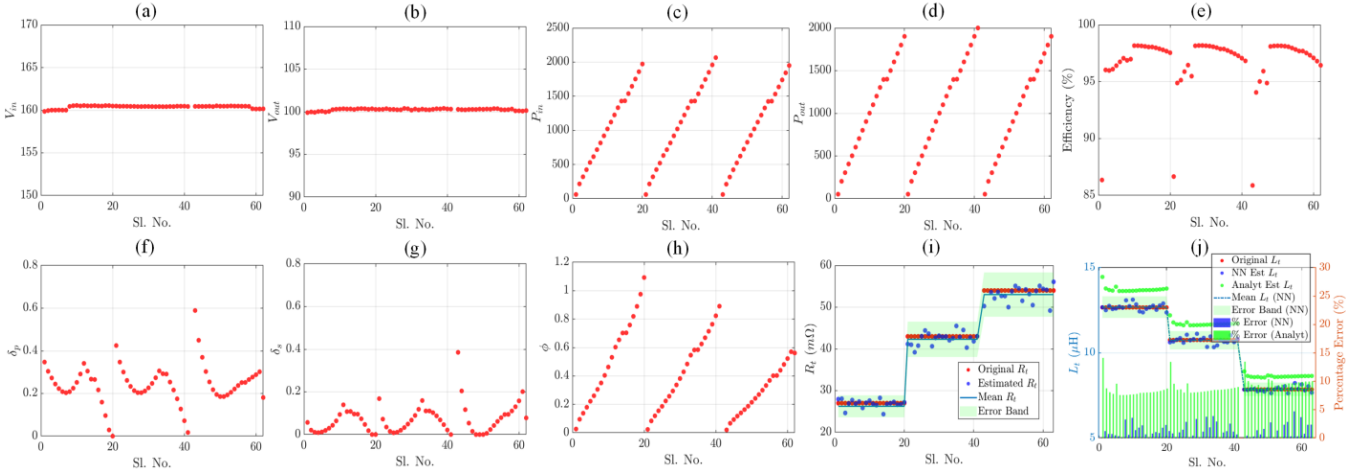

Fig. 24. NN-in-loop Performance Tracking of the Proposed DAB Modulation in hardware. Real time sensed and controller data received by NN through SPI communication includes (a) $V_{in}$, (b) $V_{out}$, (c) $P_{in}$, (d) $P_{out}$, (e) efficiency, (f) $\delta_p$, (g) $\delta_s$, (h) $\varphi$. The output data from the NN are presented in (i) $R_t$, (j) $L_t$.

TABLE VI
DESIGN SPECIFICATION OF THE FABRICATED DAB CONVERTER

| Categories of DAB Modulation Schemes | Works | $f_{sw}$ | Circuit Param Adaptive | ZVS Range | Cond. Loss | Opt. Target | Implementation Strategy | Implement. Complexity | Compute Time | Memory Requirement | High-BW Current Sensing | Loss Opt. Accuracy |
|---|---|---|---|---|---|---|---|---|---|---|---|---|
| Power-loss-model-based offline optimization | [24], [40] | fixed | N | High | Low | Total Loss | Look-up Table | Mid | Low | High | N | High |
| Inductor Current Peak optimization techniques | [17] | fixed | N | Low | Low | peak current | Closed form solution | Mid | Mid | Low | N | Low |
| Inductor Current RMS optimization techniques | [11] | fixed | Y | Low | Low | RMS current | Closed form solution | Low | Low | Low | N | Low |
| | [19], [20] | fixed | N | Low | Lowest | RMS current | Closed form solution | Mid | Mid | Low | N | Low |
| ZVS range optimization techniques | [15], [16], [21] | fixed | N | High | Mid | ZVS + peak current | Closed form solution | Mid | Mid | Low | N | Mid |
| | [14], [22] | var | N | Highest | Mid | ZVS + peak current | Look-up Table/ Closed form Sol. | Mid | Low/Mid | High/ Low | N | Mid |
| NN-based modulation | [18] | fixed | N | High | Mid | ZVS + peak current | PA-RNN based optimization | High | High | High | Y | High |
| | [25] | fixed | N | Low | Low | peak current | AI-data trimming | High | High | High | N | Mid |
| | [38] | var | N | High | Mid | Total Loss | Trained NN based optimization | High | High | High | N | High |
| | [39] | fixed | N | Low | Low | peak current | Fuzzy inference system (FIS) | High | High | High | N | Mid |
| PINN-based parameter estimation + power-loss-model-based optimization | This work | fixed | **Y** | High | Low | **Total Loss** | PINN-in-loop TPS | High | **Low-Mid** | High | N | **High** |

Finally, a detailed comparison of the proposed NN-in-loop circuit parameter-adaptive TPS modulation with the state-of-the-art DAB modulation strategies is presented in Table VI. The comparison reveals that most existing DAB modulation strategies, including NN-based optimization schemes, lack adaptability to changes in DAB circuit parameters. In contrast, the proposed technique offers parameter adaptivity while maintaining maximum efficiency tracking. The parallel execution of the trained NN-based parameter estimation and loss-optimal duty variable calculation ensures fast operation of the control loop. This eliminates computational delays that would otherwise occur if the NN were placed in the serial path of TPS variable calculation, which is traditionally done by the existing NN-based DAB modulation optimization works and hence limits the maximum operable switching frequency. Furthermore, this work demonstrates how an intelligent, algorithm-driven real-time computation can run concurrently with the primary controller using a bilateral communication link. The proposed methodology is not only efficient but also extendable to other NN-based complex control implementations, providing a scalable solution for advanced real-time intelligent control strategies.

## VII. CONCLUSIONS

This work presents a physics-informed neural network (PINN) based approach for circuit parameter estimation, its integration into loss-optimized modulation synthesis, and its hardware implementation with experimental validation. The PINN is trained using a limited dataset derived from both an experimental testbed and a precise frequency-domain converter model that includes losses and parasitics, closely replicating real hardware behavior. The method emphasizes the ease of data generation for PINN training and seamless deployment of the trained NN in conjunction with a digital controller via standard SPI communication in real converter hardware. Leveraging sampled DC voltage, current data, and steady-state control variables, the NN accurately estimates critical DAB circuit parameters—lumped series inductance and equivalent parasitic resistance—with mean absolute errors of 2.7% and 3.1%, respectively. The estimated parameters enable the proposed parameter-adaptive TPS control strategy to dynamically adjust the DAB control variables, ensuring optimal converter efficiency through enhanced soft-switching and reduced inductor RMS currents. Experimental results demonstrate that with a 30% reduction in series inductance, the proposed control strategy yields an average efficiency improvement of 1.4% across the entire operating range compared to non-adaptive modulation. Additionally, the approach lays the groundwork for future exploration, including the use of pretrained ANN input/output mappings that can be refined via online learning, as well as extensions to other converters with a broader set of parameter predictive tasks.

## ACKNOWLEDGEMENT

The authors would like to thank Sanat Agrawal for his valuable assistance in constructing the preliminary neural network modeling and training, which contributed significantly to the development of this work.